\documentclass[pdflatex,sn-mathphys]{sn-jnl}
\usepackage[T1]{fontenc}
\usepackage[latin9]{inputenc}
\usepackage{amsmath}
\usepackage{graphicx}
\usepackage{esint}
\usepackage{lipsum}
\usepackage{bbold}

\makeatletter
\usepackage{graphicx}
\usepackage{color}
\usepackage{amsmath}
\usepackage{psfrag}
\usepackage[abs]{overpic}
\usepackage{enumitem}
\usepackage{natbib}
\usepackage{float}
\usepackage{verbatim}
\usepackage[normalem]{ulem}

\newcommand{\SM}[1]{the Supplementary Material}

\newcommand{\wN}{\widehat{N}}
\newcommand{\wphi}{\widehat{\varphi}}

\definecolor{JuelichColor}{HTML}{023d6b}
\definecolor{JuelichColor2}{HTML}{98b9e2}
\definecolor{JuelichColor3}{HTML}{4D7BA7}
\definecolor{SoftAlertRed}{HTML}{AC2530}
\definecolor{SoftAlertRed2}{HTML}{90293A}

\makeatother

\usepackage{babel}
\begin{document}

\title{Compact description of quantum phase slip junctions}

\author*[1]{\fnm{Christina} \sur{Koliofoti}}\email{c.koliofoti@fz-juelich.de}

\author[1]{\fnm{Roman-Pascal} \sur{Riwar}}\email{r.riwar@fz-juelich.de}

\affil*[1]{\orgdiv{Peter Gr\"unberg Institute, Theoretical Nanoelectronics}, \orgname{Forschungszentrum J\"ulich}, \orgaddress{\city{J\"ulich}, \postcode{D-52425 }, \country{Germany}}}

\abstract{
Quantum circuit theory is a powerful and ever-evolving tool to predict the dynamics of superconducting circuits. In its language, quantum phase slips (QPSs) are famously considered to be the exact dual to the Josephson effect. However, this duality renders the integration of QPS junctions into a unified theoretical framework very difficult, and as we show, gives rise to serious inconsistencies for different formalisms, and in some cases difficulties to include time-dependent flux driving. We propose to resolve these issues by reducing and compactifying the Hilbert space describing the QPS processes. Our treatment provides for the first time a unified description of the Aharonov-Bohm and Aharonov-Casher effects, properly defines the valid form of inductive interactions to an environment, and allows to account for recent insights on how to include electromotive forces. Finally, we show that the compactification is likewise important for correctly predicting the available computational space for qubit architectures involving QPS junctions.
}

\maketitle

\section{Introduction} 

Given the enormous potential of superconducting circuits for realizing large scale quantum computers~\cite{Arute_2019,IBM_roadmap}, it is of utmost importance to provide a concise, yet powerful tool for their theoretical description. The paradigm of circuit quantum electrodynamics (cQED)~\cite{Yurke_1984,Devoret_1997,Burkard_2004,Ulrich_2016,Vool_2017} seems to provide just that: circuits are straightforwardly reduced to lumps, described by the canonically conjugate pair of total Cooper pair number $N$ and superconducting phase $\varphi$, whose operators satisfy $[\widehat{\varphi},\widehat{N}]=i$.

Despite its widespread use, quantum circuit theory is still not quite without occasional teething troubles. For instance, it was very recently argued that a proper, realistic description of circuits driven via time-varying magnetic fields requires going way beyond a simple lumped-element picture~\cite{You2019,Riwar_2022,Kenawy_2022}. One key insight of Ref.~\cite{Riwar_2022} will be of particular relevance here: for devices involving Josephson junctions, the precise form of the electromotive force is not dominated by the junction self-capacitances (as was prior consensus), but depends on the device geometry and distribution of the magnetic field.

The other central issue at the heart of this work concerns charge quantization, respectively the compactness of the superconducting phase. Charge and phase being canonically conjugate, the quantization unit of $N$ fixes the periodicity (compactness) of $\varphi$~\cite{Likharev_1985}. Importantly, whether charge should actually be quantized in general, gave rise to some buzz in recent years, with some advocating against~\cite{Thanh_2020} and others in favour of it, leading to a veto~\cite{Mizel_2020} on the predicted charge-noise insensitivity of the fluxonium~\cite{Koch_2009,Manucharyan_2009,Catelani_2011}, doubts~\cite{Murani_2020,Hakonen_2021,Murani_2021} regarding the existence of the dissipative quantum phase transition in Josephson junctions (JJ)~\cite{Schmid_1983,Bulgadaev_1984,Guinea_1985,Schoen_1990,Ingold_1999}, or relatedly, limitations on the validity of the spin-boson paradigm~\cite{Kaur_2021}. Moreover, in \textit{transport} topological phase transitions~\cite{Riwar2016,Yokoyama_2015,Strambini_2016,Vischi_2016,Eriksson2017,Yokoyama:2017aa,Repin_2020,Fatemi_2020,Peyruchat_2020,Klees_2021,Weisbrich_2021,Herrig_2022}, the compactness of the superconducting phase guarantees the conservation of topological charges~\cite{Riwar2016} via the Nielsen-Ninomiya theorem~\cite{Nielsen_1981}.
This overall flurry prompted the proposition~\cite{Roman2021} that charge quantization depends ultimately on the spatial resolution with which charge is measured or interacted with (similar in spirit to a recent pedagogical review~\cite{Devoret_2021}), but that it is nonetheless relevant in many cases to have effective low-energy theories underpinned by a representation where the phase can be compact.

In this work, we take the above new developments and persisting controversies as a context to revisit a particularly important and widely studied type of circuit element: the quantum phase slip (QPS) junction~\cite{Giordano1988,Bezryadin2000,Lau_2001,Buchler2004,Mooij_2005,Mooij_2006,Arutyunov2008,Astafiev_2012,deGraaf_2018,Li_2019,Shaikhaidarov_2022}. In their influential work, Mooij and Nazarov~\cite{Mooij_2006} put forth the idea that QPS junctions can be described as a nonlinear capacitor $\sim \cos(2\pi N)$ and thus be considered as exact duals of regular JJs, $\sim\cos(\varphi)$. QPS junctions are now an integral part of the ``zoo'' of elements which may enter any quantum circuit diagram. Their understanding as a nonlinear capacitor was used to explain observed interference patterns with respect to applied gate voltages, interpreted as the circuit version of the Aharonov-Casher effect, a recurrent and important theme in superconducting circuits~\cite{Pop2012,Manucharyan_2012,deGraaf_2018}. Very recently, it gave rise to the prediction, that the Gottesman-Kitaev-Preskill (GKP) code~\cite{Gottesman_2001} may be realized using only transport degrees of freedom~\cite{Thanh2019}.

However, there are some important issues with this simple description -- some of which are known, and others that we uncover below. For instance, it is already known that one has to tread carefully when trying to include non-linear capacitors into a cQED framework. Its standard formulation~\cite{Devoret_1997,Burkard_2004,Vool_2017} is based on distinguishing between inductive and capacitive elements, whose respective energy contributions to the Lagrangian are treated as potential and kinetic energy terms, respectively. However, nonlinear capacitors give generally rise to nonconvex kinetic energy functions with non-invertible charge-voltage relationships. To cure such problems, an alternative formulation was given in Ref.~\cite{Ulrich_2016}, based on loop charges (time-integral of loop currents). Here, the roles of inductive and capacitive elements are reversed (the former now being of kinetic nature). In alignment with the terminology of Ref.~\cite{Ulrich_2016}, we refer to these two approaches as the standard node-flux and the standard loop-charge formalisms.

In our work, we first of all show that a purely inductive treatment for QPS junctions is possible, making it compatible with node-flux quantization. However, we show that the resulting standard node-flux treatment of QPS junctions is fundamentally inconsistent with the loop-charge version already for very simple model circuits. 
In particular, it seems as if none of the approaches is fully able to account for both the Aharonov-Bohm and Aharonov-Casher effects. Moreover, the two formalisms allow for very different inductive couplings to an electromagnetic environment. In particular, we show that the standard node-flux approach would allow for a parameter regime, where QPS junctions could exhibit dissipative quantum phase transitions, which were recently put into question by Ref.~\cite{Murani_2020}. The loop-charge formalism does not provide such a regime. Finally, we find that specifically the loop-charge formalism struggles with taking into account time-dependent flux drive consistent with the recent findings of Ref.~\cite{Riwar_2022}, as it requires a nonzero self-capacitance to include Josephson junctions.
We then show that a straightforward topological compactification of the Hilbert space describing the QPS junction can resolve the above inconsistencies. This compactification gives rise to an additional constraint on the wave function, and eliminates spurious degrees of freedom. We are thus able to present a theory capable of reproducing both the Aharonov-Casher and Aharonov-Bohm effects within a single formalism. Our proposed treatment provides a crucial constraint on inductive interaction terms, and is able to incorporate time-dependent flux drive according to Ref.~\cite{Riwar_2022}. Finally, the compactification also provides important information on the available computational space for quantum information applications. Thus, we identify important caveats for the proposition by Ref.~\cite{Thanh2019} to use QPS junctions for a realization of the GKP code.

Paradoxically, the constraint on the Hilbert space can be shown to be related to the presence of charge quantization, in spite of the fact, that the low-energy description of the QPS junction does no longer contain a well-defined, quantized charge operator. This is therefore a peculiar manifestation of the principles derived in Ref.~\cite{Roman2021}. As pointed out in the outlook, our approach is likely the starting point for an entire series of revisions on the subject of QPS physics.

\section{Results}

\subsection{Symmetry or reduced Hilbert space?}

In fact, before talking about actual QPS junctions, let us outline one of the central issues with a much simpler (and probably less contested) example: a capacitively coupled Josephson junction with an applied gate, giving rise to the well-known charge qubit Hamiltonian~\cite{Cottet_2002},
\begin{equation}\label{eq_H_Josephson_regular}
    H_{C,J}=E_C (\wN+N_g)^2-E_J \cos(\wphi)\ ,
\end{equation}
with $[\wN,\wphi]=i$. The capacitive energy is $E_C=2e^2/C_\text{tot}$ ($C_\text{tot}=C+C_g$, with the junction self-capacitance $C$ and the gate capacitance $C_g$) and the Josephson energy $E_J=I_c/2e$ (where $\hbar=1$), proportional to the junction's critical current $I_c$. The parameter $N_g$ represents the gate-induced offset charge on the island, $N_g=C_gV_g/2e$, where $V_g$ is the applied voltage.

We detect right away the symmetry $H_{C,J}(\varphi+2\pi)=H_{C,J}(\varphi)$, which is well-known to express the fact that the junction transports Cooper-pairs in integer portions. But if this discrete symmetry was the only ingredient, we would find (according to Bloch's theorem) continuous energy bands with a wave vector $k$ as a quantum number, that is, $H_{C,J}\vert\psi_n\rangle=E_n(k)\vert\psi_n\rangle$. Related to that, any stationary contribution to the gate-induced offset charge $N_g$ could be gauged away by means of a time-independent unitary transformation. In order to correctly predict the experimentally well-established $N_g$-dependence~\cite{Bouchiat_1998,Nakamura_1999,Serniak2018,Serniak2019}, and discrete energies instead of energy bands, one needs to impose an additional symmetry constraint on the wave function itself, $\vert\psi(\varphi+2\pi)\rangle=\vert\psi(\varphi)\rangle$; with this constraint we obtain the eigenenergies $E_n(N_g)$ instead of $E_n(k)$~\cite{Cottet_2002}. This can be understood as an ``inverted'' Bloch theorem. The additional constraint on the wave function in some sense selects out the $k$ vector which is consistent with having $N_g$ as the equivalent of the vector potential. The latter is no longer a gauge degree of freedom: when progressing in $\varphi$-space by $2\pi$ the system returns to the same state (because $\varphi$ now lives on a compact circle), and thus self-interferes while picking up the phase $e^{i2\pi N_g}$.

While the above story seems straightforward, there are some subtleties with rather far-reaching consequences. Ultimately, everything revolves around the question of what the importance and the physical interpretation of the constraint on the wave function really is. Reference~\cite{Ulrich_2016} argues that the discrete symmetry in the Hamiltonian and the symmetry constraint on the wave function can be considered equivalent. We can fix a given Bloch vector at some initial time $t_0\rightarrow-\infty$, which must stay the same throughout the whole time-evolution. Thus, one could in principle identify the constant $k$ vector as the external parameter $N_g$ without having to explicitly impose boundary conditions on the wave function. This is certainly true, if the Hamiltonian keeps the discrete symmetry for all times. But what guarantees us that this symmetry has to be preserved? Take a seemingly harmless addition to the above Hamiltonian $H_{C,J}\rightarrow H_{C,J}+H_\text{ind}$, an inductive coupling of the form,
\begin{equation}\label{eq_inductive_interaction}
    H_\text{ind}\sim \wphi\cdot \wphi_\text{ext}\ .
\end{equation}
$\wphi_\text{ext}$ may represent a classical, externally applied field $\wphi_\text{ext}\rightarrow\varphi_\text{ext}(t)$, or express the inductive coupling to an electromagnetic environment \`{a} la Ingold and Nazarov~\cite{Ingold:1992aa}, with $\wphi_\text{ext}\sim\sum_j \wphi_j$, where $\wphi_j$ is the phase operator of the $j$-th LC resonator modelling the environment. Such a coupling immediately breaks the $2\pi$-periodicity of the system, and elevates the Bloch vector $k$ from a mere artefact to an actual, physically relevant quantity. Indeed, the model of a Josephson junction inductively coupled to an LC-resonator bath (via exactly such a $V$-term) was intensively studied, and predicted to give rise to a quantum-dissipative phase transition (of Schmid-Bulgadaev type)~\cite{Schmid_1983,Bulgadaev_1984,Guinea_1985,Schoen_1990,Ingold_1999}. Only very recently, doubt was cast on that paradigm~\cite{Murani_2020}, relying on exactly the premise that the inductive interaction term in Eq.~\eqref{eq_inductive_interaction} violates the above symmetry. The authors of Ref.~\cite{Murani_2020} proposed instead a model with a capacitive coupling $\sim \wN \wN_j$ preserving charge quantization. By means of a unitary, this coupling can be easily translated to an equivalent inductive interaction of the form $\sim \sin(\wphi)\wphi_j$ (i.e., the product of the Josephson junction and LC currents), which likewise preserves a $2\pi$-periodic $\varphi$. Note that this result was not received without controversy~\cite{Hakonen_2021,Murani_2021}. There is however one central criticism in Ref.~\cite{Murani_2020}, with which we agree, and formulate in our own words as follows. Discrete symmetries in the Hamiltonian are undeniably an important property for understanding the time-evolution of a closed system; but as such, they do not limit the available Hilbert space, to which an external drive, or an environment could potentially couple to -- or which could be used for storing and manipulating quantum information. Constraints on the wave function however do: they are the ultimate guiding principle to weed out illegal interaction terms and define the size of the available computational space for a qubit. 

While a $2\pi$-periodic constraint in $\varphi$-space (guaranteeing charge quantization) is fairly straightforward to defend for regular Josephson junctions, finding a similar principle for QPS junctions is considerably harder, or may even be believed to be impossible. In particular, consider that QPS junctions can be thought of as an extension of the regular linear inductor, described by an energy term $\sim\varphi^2$, which gives rise to a continuous quasicharge~\cite{Koch_2009}. Indeed, why and in what form should charge quantization here still matter -- and what possible objection could one still have against interactions like the inductive coupling given in Eq.~\eqref{eq_inductive_interaction}? This is a highly contested question, where we do not detect any consensus in the existing literature. In Ref.~\cite{Thanh2019,Thanh_2020} there emerge similar Bloch vectors that are taken very seriously by the authors, and prove to be an integral part of the predicted physics. Reference~\cite{Murani_2020} on the other hand already speculated in their paper on regular Josephson junctions, that a revision of the existing treatment of QPS processes may potentially be necessary. In this work, we provide a first step towards that goal. In the following, we first consider existing standard treatments of QPS junctions for a few examples of circuit models, and identify (as a matter of fact: incompatible) Bloch wave vectors. We then rederive the low-energy physics of QPS junctions. The outcome of that investigation is a well-defined constraint on the wave function which eliminates the Bloch wave vectors, readily applicable for arbitrary circuit models. As we show, this resolves remaining inconsistencies, and provides for the first time a clean theoretical treatment of QPS junctions compatible with the Aharonov-Bohm and Aharonov-Casher effects. Finally, we give an example where determining the correct size of the Hilbert space is crucial not only for appropriately accounting for a coupling to an external field, but to correctly assess the available state space for qubit operations.

\subsection{Node-flux versus loop-charge quantization}

Mooij and Nazarov~\cite{Mooij_2006} made an important and insightful simplification in describing the low-energy QPS physics, by reducing the Hamiltonian of a QPS wire to the form
\begin{equation}\label{eq_H_standard_QPS_alone}
    H_\text{QPS}=-\frac{E_S}{2}\sum_f\Big(\vert f\rangle\langle f-1\vert+\vert f-1\rangle\langle f \vert \Big)+E_L \sum_f\Big(\varphi+2\pi f \vert f\rangle\langle f \vert\Big)^2\ ,
\end{equation}
where $\varphi$ is the phase difference applied across the wire and $f$ counts the number of phase slips. Quantum jumps in $f$ are associated to the energy scale $E_S$. Changes in $f$ also come with an energy penalty parametrized by the inductive energy $E_L$. 

\begin{figure}[htbp]
    \begin{center}
        \includegraphics[width=0.65\textwidth]{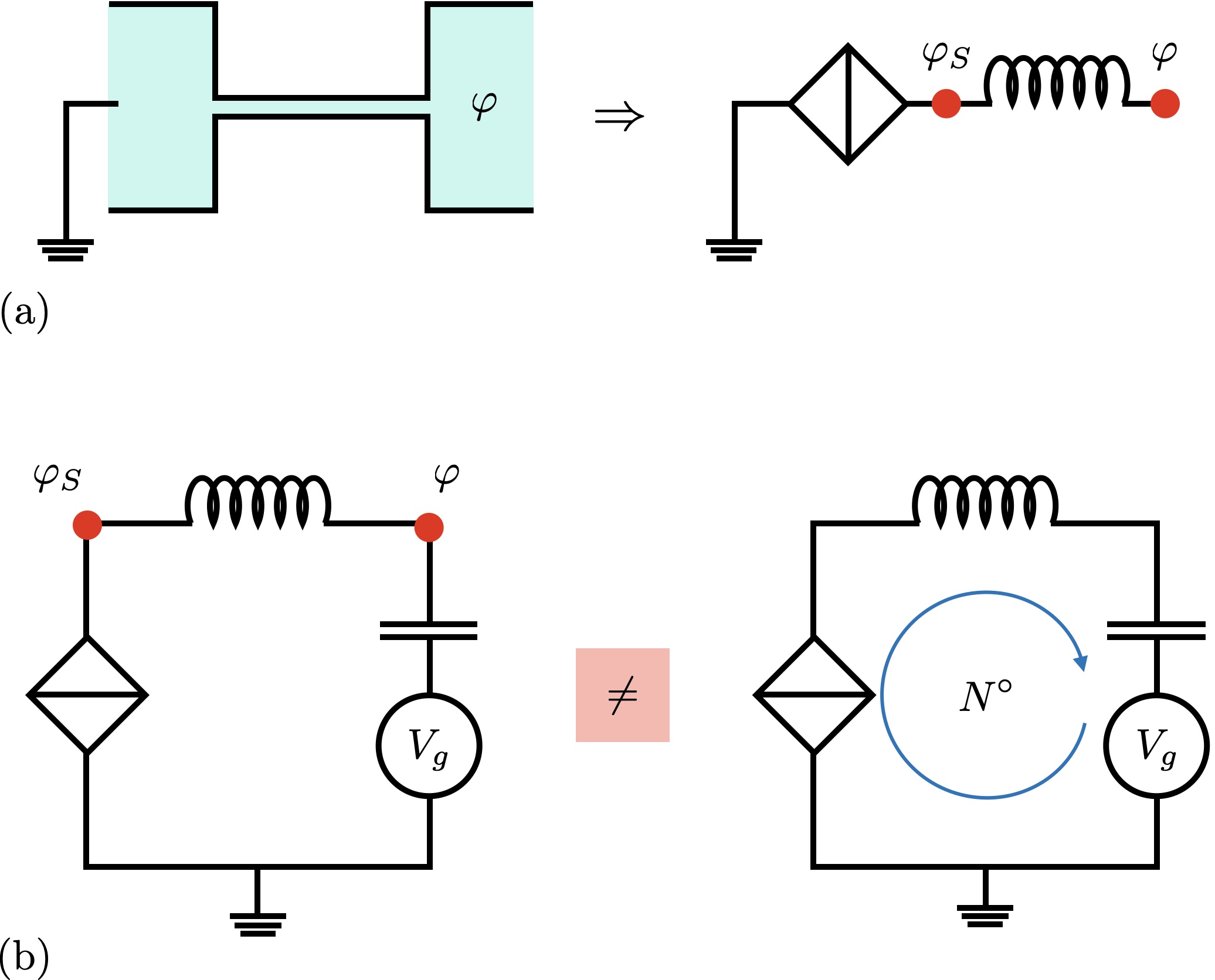}
    \end{center}
    \caption{Different flavours of standard treatments of QPS junctions and their inconsistencies. (a) A thin superconducting wire connecting two superconducting lumps (contacts) with phase difference $\varphi$ realizing a QPS junction (left), and the equivalent circuit proposed by Mooij and Nazarov~\cite{Mooij_2006}, adding an auxiliary node with phase $\varphi_S$ (right). (b) Simple circuit coupling a QPS junction with a capacitance and a gate voltage $V_g$, either treated with the ordinary node-flux quantization (left) or the dual loop charge quantization proposed in Ref.~\cite{Ulrich_2016} (right). The $\neq$-sign indicates that they disagree, as argued in the main text. 
    }
	\label{fig:QPS_standard_treatment_start}
\end{figure}

Once brought into this form, they noticed the beautiful correspondence between the above Hamiltonian and the Hamiltonian of a Josephson junction with a capacitance, see Eq.~\eqref{eq_H_Josephson_regular}. For this purpose, simply replace $E_S$ with the Josephson energy $E_J$, $4\pi^2E_L$ with the capacitive energy $E_C$, $\varphi/2\pi$ with the gate-induced offset charge $N_g$, and the number of phase slips $f$ with the number of Cooper pairs $N$ transported across the junction. Hence, the picture of quantum phase slips as the exact dual to the Josephson effect was born. The Hamiltonian was cast into the compellingly elegant form
\begin{equation}\label{eq_H_standard_QPS_alone_nonlinear_capacitor}
    H_\text{QPS}=-E_S\cos(2\pi\widehat{N}_S)+E_L (\varphi+\widehat{\varphi}_S)^2\ ,
\end{equation}
adding the auxiliary node described by the charge and phase operators $\widehat{N}_S,\widehat{\varphi}_S$ (see Fig.~\ref{fig:QPS_standard_treatment_start}a), satisfying the ordinary charge-phase commutation relation. $\varphi$ is as of right now a constant parameter. When integrated into a larger circuit (see in a moment), it will be likewise treated as a dynamical quantum variable $\varphi\rightarrow \wphi$. At any rate, Eq.~\eqref{eq_H_standard_QPS_alone_nonlinear_capacitor} allows for the interpretation of the QPS junction as a nonlinear capacitor element.

Let us at this point already interject and foreshadow the second part of this work: as we show later, when carefully revisiting the low-energy dynamics of QPS junctions, we actually already disagree with the form of the first Hamiltonian, Eq.~\eqref{eq_H_standard_QPS_alone}, on a tiny but important detail. Namely, it is commonly accepted that the phase observable can assume any value, whereas we argue in favour of a compact phase, $\varphi\in[-\pi,\pi)$, with (to be determined) periodic boundary conditions. This will be the central compactification procedure.

But for now we postpone this discussion, and accept an unbounded $\varphi$ -- and review the issues it ensues. As a matter of fact, let us first notice a couple of pitfalls which have nothing to do with whether $\varphi$ should be compact or not. In Eq.~\eqref{eq_H_standard_QPS_alone} the Hilbert space consists of one continuous~\footnote{We stress that one should read the word ``continuous'' like a physicist, and not like a mathematician. Just like the continuous eigenspectrum of a solid state is never truly continuous but only quasicontinuous due to a finite size of the system, circuit degrees of freedom are likewise subject to cut-offs. For instance, the charge on the island of a transistor can never actually go to $\pm\infty$, as we would eventually break the device. Consequently, the conjugate phase is at most quasicontinuous.} ($\varphi$) and one discrete ($f$) observable. Equation~\eqref{eq_H_standard_QPS_alone_nonlinear_capacitor} invites the wrong impression that there are two continuous observables, $\varphi,\varphi_S$, thus massively overcounting the number of states. This might seem like a minor hiccup, as one may just carry the larger number of states around, and reduce it again at an opportune moment by imposing an appropriate condition on the wave function (which is, in this case, obvious). But as we will show, mistakes are easily made. For instance, one could be too quick and take the limit $E_L\rightarrow\infty$, such that the inductive term acts as a Lagrange multiplier, perfectly coupling $\varphi$ with $\varphi_S$. This would promote the auxiliary charge $N_S$ appearing in the $E_S$ term to the physical charge $N$. This is often considered to be the limit of an ``ideal'' QPS junction. As we argue at the very end of this work, this limit does not function like that.

The second problem is that the $\cos(2\pi N)$-representation of the QPS junction is incompatible with the ordinary node-flux quantization procedure~\cite{Devoret_1997,Burkard_2004,Vool_2017}. Here, the circuit Lagrangian $L$ is constructed by adding the energies of capacitive elements to the kinetic energy, and inductive elements to the potential energy. If we interpret QPS processes as a capacitive coupling, the $E_S$-term needs to be treated like a kinetic energy. We thus have to find an energy expression depending on the phase velocity $\dot{\varphi}$, which, after the Legendre transformation $H=\dot{\varphi}N-L$ with $N=\partial_{\dot{\varphi}} L$, reduces to the $\cos(2\pi N)$ function. We can already see by the definition $N=\partial_{\dot{\varphi}} L$ that the inverse of the $\cos$ (or $\sin$) function will be involved, which is obviously not defined on the entire relevant domain of $N$, and therefore has to be treated as a multivalued function. This is clearly outside the regime of validity for the Legendre transformation (which is only defined for convex functions). For completeness, we note that such a kinetic energy function for the QPS term was nonetheless proposed in Ref.~\cite{Thanh_2020}. However, the issues of this function, in particular an uninvertible relationship between $N$ and $\dot{\varphi}$, can only be bypassed for some of the simple circuits considered in this work, and quickly get out of hand for more generic models. In fact, the authors of Ref.~\cite{Thanh_2020} provide an example themselves, when connecting an ``ideal'' QPS junction and a regular linear capacitance.

Highly alert of these issues, Ulrich and Hassler~\cite{Ulrich_2016} proposed a dual quantization procedure based on loop charges (i.e., time-integrals of loop currents). Here, the roles of inductive and capacitive elements is reversed: the former are now kinetic energy terms and the latter are added to the potential energy. Thus, QPS junction elements (now part of the potential energy) allow for the Legendre transformation to be performed without a hitch. This approach however comes with its own set of issues, as we show throughout this section. 

Some of these issues (especially related to time-dependent flux drive) will provide a strong incentive to nonetheless try and remain in the regular node-flux formalism. We argue here that this is possible -- a first central result of our work. To this end, notice that while $N_S$ seems to take the form of a charge, it is really an auxiliary charge representing the QPS processes, and there is so far no indication that it can be interpreted in any meaningful way as a literal, physical charge (we will develop a more precise interpretation in the next section). And as such it can only couple to the rest of the circuit through the $E_S$ and $E_L$ energy terms in Eq.~\eqref{eq_H_standard_QPS_alone_nonlinear_capacitor}. For instance, we do not deem it physical to all of a sudden couple the auxiliary charge to ground via an additional capacitance. Then, we can in principle circumvent the problems related to the non-linear capacitor treatment of the QPS junction, by simply not treating it as a capacitor at all. Instead of seeing Eq.~\eqref{eq_H_standard_QPS_alone_nonlinear_capacitor} as a sum of a capacitive and an inductive term, which have to enter the Lagrangian in different places (either in the kinetic or potential energies), we treat the \textit{entirety} of the Hamiltonian as an inductive term. Then, there is also no more need nor formal advantage for the interpretation of the internal QPS degree of freedom as auxiliary charge and phase, $N_S$ and $\varphi_S$. We can therefore return to Eq.~\eqref{eq_H_standard_QPS_alone} which is devoid of such interpretations. Here, $f$ is just the number of phase slips, nothing more, nothing less. We then add this Hamiltonian in its totality to the potential energy of the Lagrangian for the conventional node-flux formalism. As a consequence, it will of course not just be a scalar potential energy, but an operator. But this does not provide any formal obstacle, as neither the Legendre transformation, nor the charge-phase quantization procedure are really affected by it. Note that similar strategies have been adopted to incorporate the fractional Josephson effect of Majorana-based junctions (which is likewise no longer a scalar energy term) into larger circuit models~\cite{van_Heck_2012}.

Only in the very last step, once we have already safely performed the Legendre transformation, may we reinsert the auxiliary charge and phase operators $\wN_S,\wphi_S$ into the Hamiltonian if we so wish for aesthetic reasons. In what follows, we adopt this strategy, and (even though it has not been explicitly proposed yet) call it the standard node-flux approach -- to contrast it to the later developed compact node-flux approach, where we implement the foreshadowed compactification of $\varphi$. 
In the remainder of this section, we compare the resulting Hamiltonians obtained by standard node-flux and loop-charge quantization for chosen examples of circuit models.

\subsection{Issues and inconsistencies}

First, we consider a device where one end of the QPS junction is grounded, and the other is capacitively coupled to a gate voltage, see Fig.~\ref{fig:QPS_standard_treatment_start}b. The regular node-flux quantization procedure (treating the entire QPS junction as an inductive element as introduced above) provides the Hamiltonian
\begin{equation}\label{eq_H1QPS_node}
    H_{C,\text{QPS}}=E_C(\wN+N_g)^2-E_S\cos(2\pi\wN_S)+E_L(\wphi+\wphi_S)^2\ ,
\end{equation}
where here, $E_C=2e^2/C_g$ and $N_g=C_g V_g/2e$. In contrast, applying the rules for loop-charge quantization (for details, see Ref.~\cite{Ulrich_2016} and \SM{}), we get the Hamiltonian
\begin{equation}\label{eq_H1QPS_loop}
    H_{C,\text{QPS}}^\circ=E_C(\wN^\circ+N_g)^2-E_S\cos(2\pi\wN^\circ)+E_L(\wphi^\circ)^2\ ,
\end{equation}
where loop charge and loop flux likewise satisfy $[\wN^\circ,\wphi^\circ]=i$. We use the notation $X^\circ$ for quantitites $X$ that are specific to the loop-charge quantization procedure.

The two formalisms do not predict the same number of degrees of freedom, one loop versus two nodes. Inspired by Ref.~\cite{Ulrich_2016}, we might try to render the two approaches equivalent by looking for symmetries. Already here, there is a first trap: one might naively identify a continuous symmetry when translating $\wphi\rightarrow \wphi+\delta\varphi$ and $\wphi_S\rightarrow \wphi_S-\delta\varphi$. Consequently, as per Noether's theorem, we would find the conserved quantity $\wN-\wN_S$ commuting with the Hamiltonian, which could thus be eliminated from the dynamics. However, remember that the auxiliary phase $\varphi_S$ originally came from the discrete number of phase slips $f$ in Eq.~\eqref{eq_H_standard_QPS_alone}. Therefore, crucially, the above identified symmetry cannot possibly be \textit{continuous}, but must be \textit{discrete}. Noether's theorem does not apply -- but Bloch's theorem does. The analysis thus all of a sudden strongly resembles the initial discussion we had regarding the Josephson junction. Indeed, we could eliminate the resulting Bloch wave vector $k$ in the node-flux Hamiltonian, if there was a way to impose the discrete symmetry in ($\varphi$,$f$)-space onto the wave function. This would not only equalize the number of degrees of freedom, but it would also make the predictions consistent with respect to $N_g$. Namely, the discrete energy levels of the loop-charge Hamiltonian are $N_g$-dependent (in Eq.~\eqref{eq_H1QPS_loop} $N_g$ cannot be removed), whereas in Eq.~\eqref{eq_H1QPS_node} $N_g$ is a mere gauge term. But as is the central theme of this work, such a constraint on the wave function needs to be justified.

A highly related, and equally important consequence can be best appreciated when performing a simplification of the QPS term for $E_S\gg E_L$. In this parameter regime we can approximate the effect of the QPS wire by what looks like an effective Josephson junction, with Josephson energy
\begin{equation}\label{eq_EJeff}
    E_{J,\text{eff}}\approx 16 \sqrt[4]{\frac{E_L E_S^3}{2}}e^{-\sqrt{\frac{8 E_S}{\pi^2 E_L}}}\ .
\end{equation}
This approximation of the QPS junction is valid as long as the linear capacitor does not excite it, $E_C<\sqrt{E_S E_L}$. Importantly however,  the standard node-flux formalism does not provide an exact mapping to a Josephson junction, because there is no periodic constraint on $\varphi$. Therefore, the formalism does in principle no longer prevent the existence of an inductive coupling term as given in Eq.~\eqref{eq_inductive_interaction}. The quantum-dissipative phase transition put into question in Ref.~\cite{Murani_2020} for regular Josephson junctions would all of a sudden be back on the table for QPS junctions. Importantly, the same is not true for the loop-charge version of the Hamiltonian. An interaction term of the form in Eq.~\eqref{eq_inductive_interaction} would here be possible by replacing $\wphi$ with $\wphi^\circ$. We stress however, that is \textit{not} the same interaction, because $\wphi$ and $\wphi^\circ$ are not the same operators (as we show in the second part, $\wphi^\circ$ is mapped to $\wphi+2\pi \widehat{f}$, with compact $\varphi$). In particular, the resulting inductive interaction after the approximation $E_S\gg E_L$ could not possibly break the newly gained symmetry of the effective Josephson effect, as (contrary to the node-flux approach) there is simply not enough available Hilbert space.


\begin{figure}[htbp]
    \begin{center}
        \includegraphics[width=0.65\textwidth]{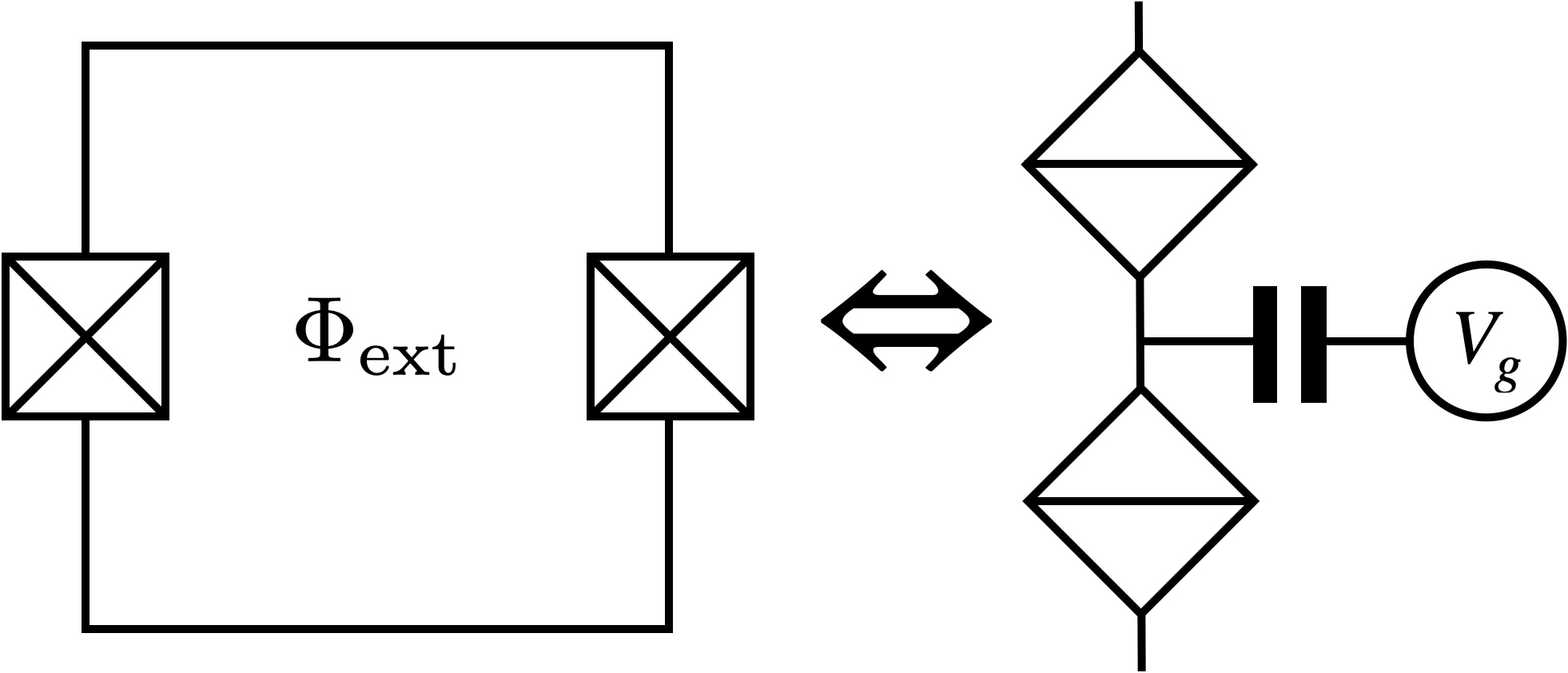}
    \end{center}
    \caption{Aharonov-Bohm effect versus Aharonov-Casher effect in circuit-QED. While the Aharonov-Bohm effect is present and measurable as a sensitivity of the dc-SQUID (left) eigenspectrum of the applied external flux $\Phi_\text{ext}$, the experimentally observed $N_g$-dependence of a circuit connecting two QPS junctions in series (right) is the dual Aharonov-Casher effect. In this work, we provide an alternative understanding for the latter. }
	\label{fig:Aharonov_Casher_vs_QPS}
\end{figure}

\begin{figure}[htbp]
    \begin{center}
        \includegraphics[width=0.65\textwidth]{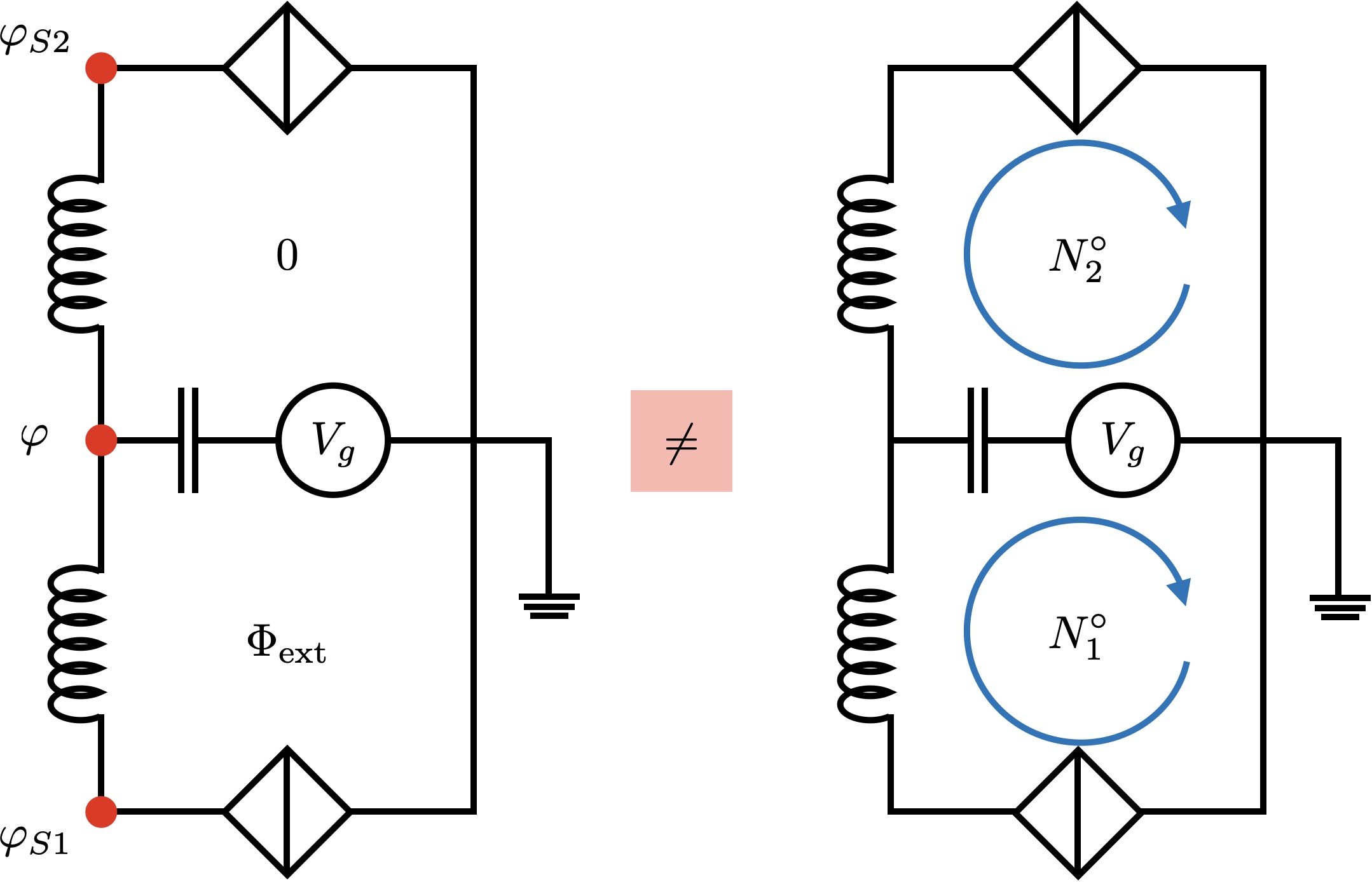}
    \end{center}
    \caption{Inconsistencies between different formalisms for a device with two QPS junctions. While the node-flux formalism predicts an Aharonov-Bohm effect, the loop-charge formalism predicts an Aharonov-Casher effect, but the two cannot coexist.}
	\label{fig:QPS_junctions_standard_description_AB_AC}
\end{figure}

In addition to external gate voltages, we now consider devices where flux-control via magnetic fields is important. In such a setup, one should generally expect an interesting interplay between the Aharonov-Bohm effect and the cQED version of the Aharonov-Casher effect, where the latter is dual to the former, see Fig.~\ref{fig:Aharonov_Casher_vs_QPS}. Consider therefore a circuit model with two QPS junctions, see Fig.~\ref{fig:QPS_junctions_standard_description_AB_AC}.
Through standard node-flux quantization, we arrive at the Hamiltonian
\begin{equation}\label{eq_H_2QPS_node}
\begin{split}
    H_{C,\text{2QPS}}=E_C(\wN+N_g)^2-E_{S1}\cos(2\pi \wN_{S1})+E_{L1}(\wphi+\wphi_{S1}+\phi_\text{ext})^2\\-E_{S2}\cos(2\pi \wN_{S2})+E_{L2}(\wphi+\wphi_{S2})^2\ .
\end{split}
\end{equation}
The standard loop-charge quantization approach on the other hand delivers (see \SM{})
\begin{equation}\label{eq_H_2QPS_loop}
\begin{split}
    H_{C,\text{2QPS}}^\circ=E_C(\wN_1^\circ-\wN_2^\circ+N_g)^2-E_{S1}\cos(2\pi \wN_{1}^\circ)+E_{L1}\wphi_{1}^{\circ 2}\\-E_{S2}\cos(2\pi \wN_{2}^\circ)+E_{L2}\wphi_{2}^{\circ 2}\ .
\end{split}
\end{equation}
Very briefly, concerning the $N_g$-dependence we do not gain any new insights. We see that in the flux-node formalism, a stationary $N_g$ is just as much a gauge degree of freedom here, as it already was with just one QPS junction, Eq.~\eqref{eq_H_standard_QPS_alone_nonlinear_capacitor}. Yet again, there is a discrete symmetry in phase space, leading to a Bloch wave vector. The spectrum of the loop-charge Hamiltonian on the other hand remains $N_g$-dependent even with the addition of the second QPS junction.

Let us now turn our attention towards the impact of $\phi_\text{ext}$. While the spectrum of the node-flux Hamiltonian $H_{C,\text{2QPS}}$ is constant in $N_g$ (as just noted), it is worth pointing out that it actually contains a nonzero $\phi_\text{ext}$-dependence. As can be seen in Eq.~\eqref{eq_H_2QPS_node}, $\phi_\text{ext}$ cannot be removed by a unitary transformation. Such a removal could only work if we could incorporate the external flux into a shift of $\varphi_{S1}$.
Once again, this would require $\varphi_{S1}$ to be a continuous variable, which it is not (as we have explained above). The loop-charge Hamiltonian has the reversed feature: its spectrum is $N_g$-dependent but constant in $\phi_\text{ext}$ (in fact, the latter does not exist in the formalism). However, similarly to the node-flux Hamiltonian, we now get a discrete symmetry (and ensuing Bloch wave vector), except that it is here in charge space instead of phase space. Therefore, both formalisms predict the existence of Bloch bands, except that the respective $k$-vectors live in opposite (dual) spaces. 

Our investigation so far reveals the following: in the absence of plausible constraints on the Hilbert space, the existing formalisms exhibit a fundamental incompatibility. In particular, it seems as if the Aharonov-Bohm effect and the Aharonov-Casher effect cannot coexist. Moreover, the two formalisms allow for very different versions of inductive coupling to an enviroment, which could potentially decide over the existence or absence of quantum-dissipative phase transitions~\cite{Murani_2020}. But we also see, that they could be rendered equivalent, if we indeed find additional constraints. As already stated, a part of the community does so far not expect a constraint in ($\varphi$,f)-space for the node-flux approach, which would be equivalent to some notion of charge quantization. As for the loop-charge treatment of the circuit with two QPS junctions, the necessary constraint would be equivalent to stating that the sum of the two loop charges $N_1^\circ+N_2^\circ$ can only assume values within an interval of length $1$ -- or in the conjugate space, there should be some notion of flux quantization. Also here, such a quantization condition must be derived and justified. Without constraints, one cannot exclude the existence of interaction terms, here, e.g., a capacitive coupling of the form $\sim \wN^\circ \cdot \wN_\text{ext}$, which would yet again break the symmetry, and render the Bloch wave vector highly physical. Within the formalism given by Ref.~\cite{Ulrich_2016}, no obvious principle prevents the addition of such interactions.

What is more, note that the two discrete symmetries for the two formalisms live in different spaces (charge or phase space, respectively). It is therefore far from obvious how both constraints could in any way be related, and unified with one single formalism. In the now following section, we show how such a unification can be accomplished. By the way, in doing so, our formalism will provide a neat alternative explanation for the Aharonov-Casher effect, based on a topological principle.

Two more remarks are in order. On the one hand, let us note that there might already exist some tentative indication for the coexistence of the Aharonov-Casher and Aharonov-Bohm effects in already published experiments studying QPS junctions in series~\cite{Astafiev_2012,deGraaf_2018}. An $N_g$-dependent spectrum was measured on both occasions, which either invalidates the standard node-flux formalism in favour of the loop-charge version, or hints at the presence of some form of charge quantization. In order to match the experimental data with a model, an empirical Hamiltonian was deployed in these works. This Hamiltonian is usually given in the literature without explicit derivation from more fundamental principles. It is therefore interesting to note that we can propose such a derivation here, by taking a special limit of the loop-charge Hamiltonian above: for $E_C\rightarrow \infty$, the capacitive energy essentially becomes a Lagrange multiplier, coupling the charges as $\wN^\circ\equiv \wN^\circ_1=\wN^\circ_2-N_g$, resulting in
\begin{equation}\label{eq_H_2QPS_loop_AC}
    H_{C,\text{2QPS}}^\circ\approx-E_{S1}\cos(2\pi \wN^\circ)-E_{S2}\cos(2\pi [\wN^\circ-N_g])+(E_{L1}+E_{L2})(\wphi^{\circ})^2 \ .
\end{equation}
We note that a similar derivation by means of the standard node-flux formalism is impossible, due to the fact that $N_g$ can be gauged away here.
Depending on the value of $N_g$ the two cosines of the QPS processes interfere either constructively or destructively. This picture of an interference of phase slip processes of neighbouring QPS junctions is interpreted as the cQED version of the Aharonov-Casher effect, dual to the flux-induced interference of a regular dc-SQUID (due to the Aharonov-Bohm effect). In the absence of flux quantization, the above Hamiltonian should provide $N_g$-dependent Bloch bands, $E_n(k,N_g)$. The aforementioned experiments however seem to detect discrete energy lines, indicative of such a constraint. Of course, by itself, this finding is not yet ironclad proof. Even if Bloch bands existed, they could be very thin, beyond the available resolution limit. More convincing proof could be achieved if a spectrum depending on both $N_g$ and $\phi_\text{ext}$ was measured. To the best of our knowledge, such experiments have not yet been conducted.

\begin{figure}[htbp]
    \begin{center}
        \includegraphics[width=0.75\textwidth]{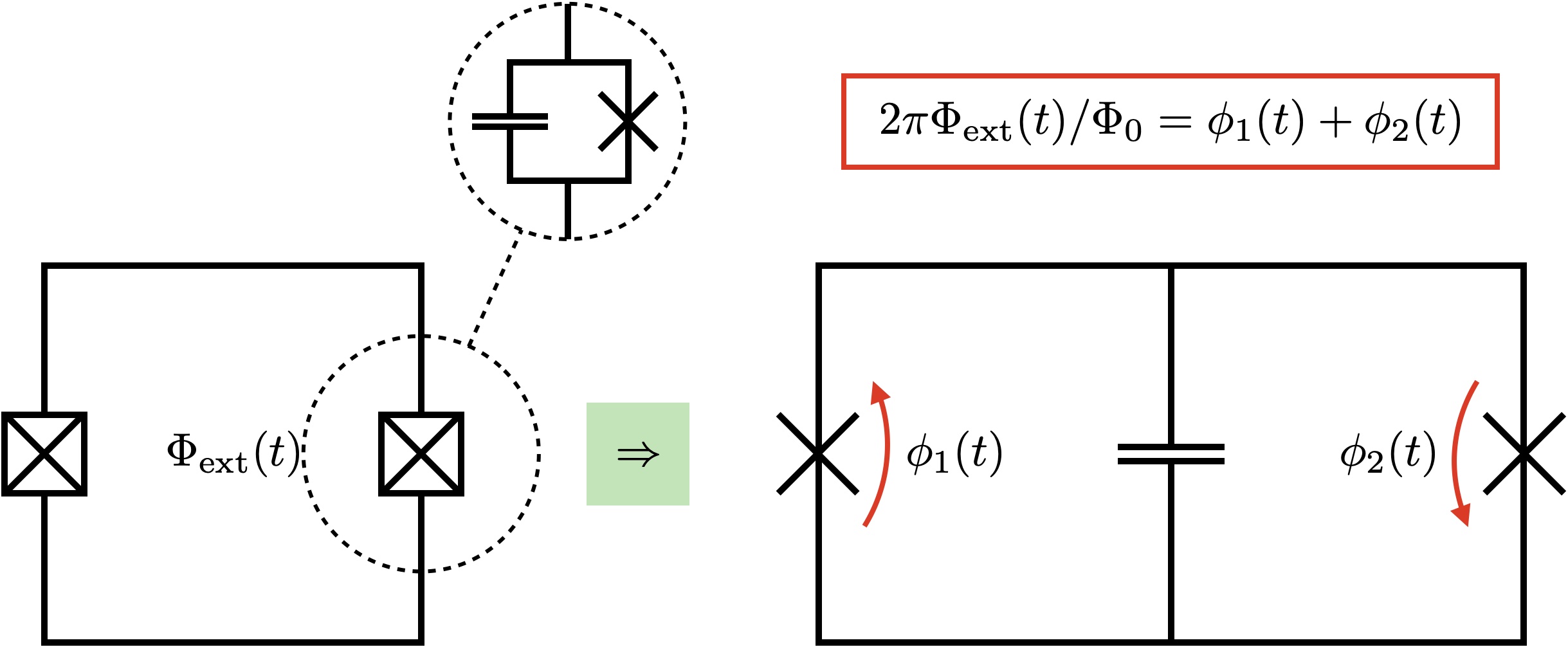}
    \end{center}
    \caption{Final issues specific to loop-charge quantization, related to Josephson junctions and time-dependent flux control. The loop-charge based quantization procedure~\cite{Ulrich_2016} can only include Josephson junctions in a mixed node/loop picture, requiring extra capacitive shunts for each junction (left). However, as shown in Ref.~\cite{Riwar_2022}, the capacitively shunted Josephson junction picture does not hold in general for time-varying magnetic fields, and external fluxes have instead to be attached to the bare junctions (right).  }
	\label{fig:QPS_junctions_standard_description_time_dependence}
\end{figure}

A second and final remark concerns a weakness specific to the loop-charge formalism, when including time-dependent flux driving in combination with Josephson junctions. In the loop-charge formalism, Josephson junctions can only be taken into account when assuming a self-capacitance shunting each Josephson junction (see Fig.~\ref{fig:QPS_junctions_standard_description_time_dependence}a), such as to accommodate a mixed loop/node picture. However, Ref.~\cite{Riwar_2022} recently argued, that the electrodynamically soundest approach is to neglect the junction self-capacitances (physically there only exists one bulk capacitance for the entire island), and add the time-dependence directly inside the Josephson junction elements, see Fig.~\ref{fig:QPS_junctions_standard_description_time_dependence}b. In order for this picture to be incorporated into the loop-charge formalism, we expect that one would have to introduce auxiliary self-capacitances, which subsequently have to be put to zero. This is not strictly speaking an insurmountable problem, but nonetheless renders the formalism unnecessarily cumbersome. In Ref.~\cite{Riwar_2022}, an alternative picture was established which could be of use for the loop-charge formalism. Namely, a generic driving can indeed be mapped back onto a circuit where the dominant capacitance contributions come from the Josephson junctions. However, in order for this mapping to work, one has to accept either partially negative, time-dependent, and even momentarily singular self-capacitances. Then, it is unclear, whether the mixed loop/node solution still works, as one might have to deal with intermediate steps where partial kinetic energies are yet again non-convex functions. At any rate, as stated, we choose to avoid these potential complications by sticking to the node-flux picture.

\subsection{Compact versus noncompact phase}

In the light of above issues, let us revisit the low-energy physics of QPS junctions and justify a version of charge quantization, which still applies for these elements. While QPS wires may be realized in a number of ways, our discussion is aimed at features which are the same, irrespective of the experimental realization.

\begin{figure}[htbp]
    \begin{center}
        \includegraphics[width=0.45\textwidth]{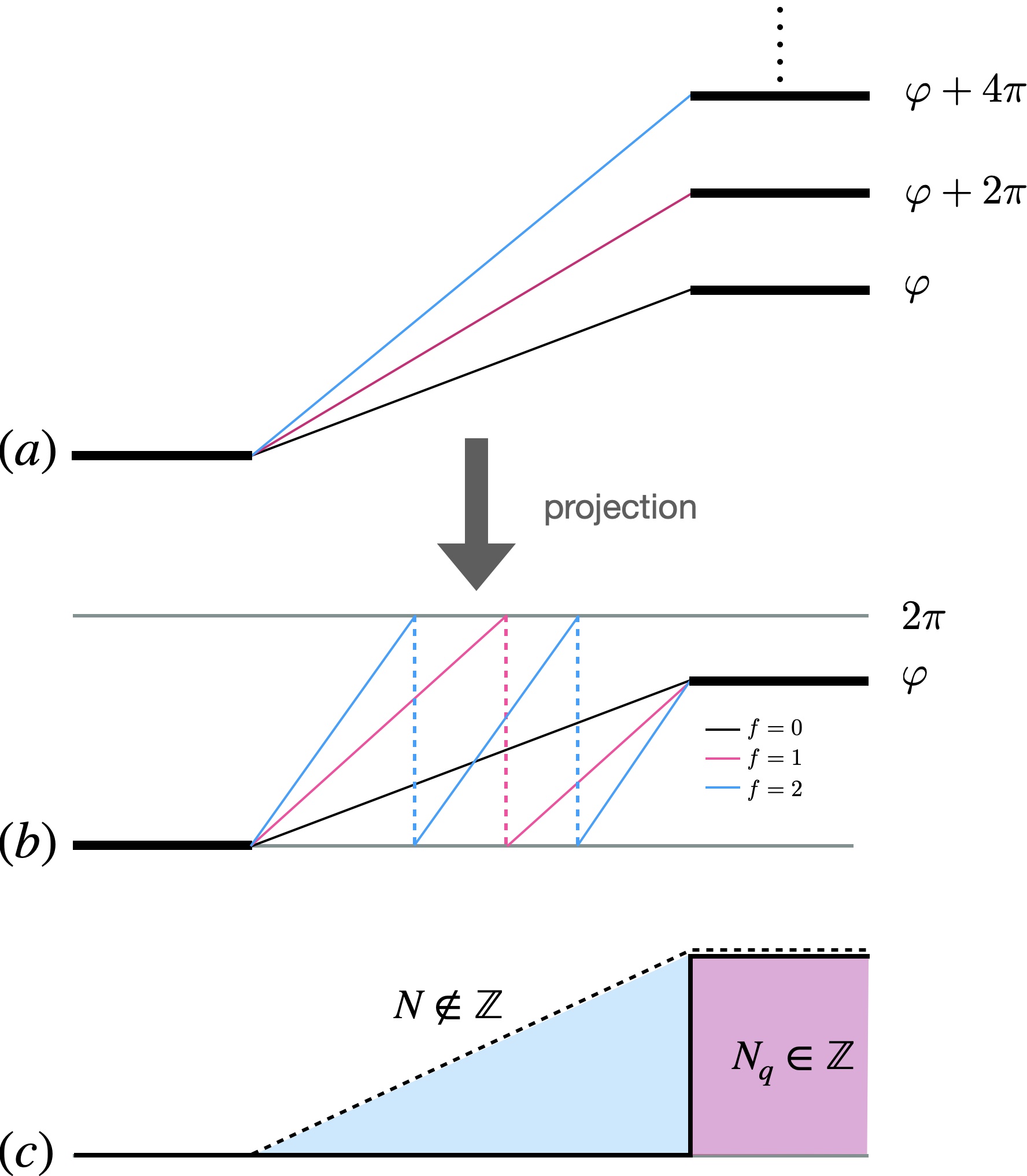}
    \end{center}
    \caption{Compact versus extended descriptions of quantum phase slips along a QPS junction (e.g., a thin superconducting wire). The two ends are phase biased by $\varphi$ (the left contact is the ground with phase 0). The extended phase profile in (a) can be projected to a $2\pi$-periodic profile in (b). While it seems that picture (a) is the one motivating the standard Hamiltonian description by Mooij and Nazarov, we argue in the main text that the standard model contains too large a Hilbert space including spurious states which are represented in neither (a) nor (b). (c) Depiction of two possible measurements of charges transported across the QPS junction, the quantized charge $N_q$ and the continuous charge $N$. While the quantized charge $N_q$ is effectively eliminated in the course of the low-energy approximation, its presence is nonetheless felt as a topological constraint on the Hilbert space. }
	\label{fig:PhaseSlipExpression2}
\end{figure}

Reconsider the isolated QPS junction, which is only contacted to two superconduting lumps. Just as in Fig.~\ref{fig:QPS_standard_treatment_start}a, we assume that one of the lumps has phase zero (ground), and the other phase $\varphi$.
In the absence of QPS processes, it is understood that the connecting wire has a spatially varying local phase profile, linearly ramping the phase up from zero to $\varphi$, see Fig.~\ref{fig:PhaseSlipExpression2}. This linear profile is the classical solution minimizing the internal energy of the wire, on top of which there may be local fluctuations. For a continuous wire, the frequency of local fluctuations can be estimated by means of the parameters of the Nambu-Goldstone mode~\cite{Altland_Simons_book}, which (in 1D) can be interpreted as a capacitive and an inductive density $c$ and $l$ of the wire. For Josephson junction arrays, this minimization works in a very similar way (see, e.g., Ref.~\cite{Catelani_2011}), where the densities $l$ and $c$ can be related to the junction energies, respectively, the capacitances in the array (mostly the capacitance to ground for sufficiently long arrays~\cite{Houzet_2019}). Neglecting local fluctuations is therefore justified when considering energies below the superconducting plasmon frequency $\omega_p\sim \sqrt{lc}d$, where $d$ is the wire length. Within the same framework, we can estimate also the energy associated to the strain due to the linear phase profile, which provides us with an inductive energy of the wire, $\sim E_L\varphi^2$, with $E_L=1/(8e^2 dl)$. The following low-energy description of QPS junctions is in particular justified when $E_L\ll \omega_p$ (we remind that $\hbar=1$).

As for quantum phase slips, they have been historically first understood as a quantum analogy of the classical, thermally activated phase-slips described by means of Ginzburg-Landau theory, see Ref.~\cite{Giordano1988} (and references therein). Within Ginzburg-Landau, the superconducting phase is by construction compact at each position in space, such that there is not one unique solution to the minimization profile of the phase inside the wire, but many distinct (in a field-theoretic sense local) minimum solutions, which can either be obtained by copying the phase $\varphi$ by multiple integers of $2\pi$ (see Fig.~\ref{fig:PhaseSlipExpression2}a) or equivalently in a compactified way with the number $f$ of kinks in the phase profile  (see Fig.~\ref{fig:PhaseSlipExpression2}b). The latter picture can be obtained from the former via a projection (taking the phase modulo $2\pi$). Note that in this particular case, this type of projection can actually be regarded as quasi-bijective, since we can undo the projection by demanding that the phase profile be continuous for the noncompact case. Alternatively, we can think of the projection as a ``rolling up'' of the extended phase onto a cylinder, where the bijective nature of the map becomes even more apparent.

In either way, in the quantum regime the system may now coherently jump between these different minima. We understand these as nonlocal quantum fluctuations (as opposed to the above local fluctuations, which we neglected). Taking $f$ as the number of kinks (or of $2\pi$ jumps) and associating to each local minimum the quantum state $\vert f\rangle$, Mooij and Nazarov~\cite{Mooij_2006} proposed the Hamiltonian description given in Eq.~\eqref{eq_H_standard_QPS_alone}, where the quantum coherent phase slips are included in the $E_S$-term.

As an aside: in the existing literature, a lot of effort went into finding accurate predictions for the value of $E_S$, especially for Josephson junction arrays~\cite{Glazman_1997,Gurarie_2004}, including off-set charge fluctuations within the array, which give rise to a temporally fluctuating value for $E_S$~\cite{Manucharyan_2012}, or in a very large array limit, where off-set charge disorder was argued to lead to strong values for $E_S$~\cite{Houzet_2019}. This is not our goal here. Instead, we focus entirely on revisiting the topology and size of the Hilbert space, while assuming $E_S$ as a given value, or a fitting parameter.

For a constant, given value of $\varphi$, there is so far no issue. Importantly however, the above introduced ingredients help us to identify the crucial detail missing in the standard Hamiltonian, Eq.~\eqref{eq_H_standard_QPS_alone}, once we start varying $\varphi$. If we allow for a Hilbert space where the discrete quantum number $f$ coexists with a noncompact $\varphi$ ($\varphi\in\mathbb{R}$), shifting $\varphi$ by $2\pi$ (and keeping $f$ constant) would actually give rise to a state distinct from the one where we leave $\varphi$ constant and instead shift $f$ by $1$. However, given the phase profile in Fig.~\ref{fig:PhaseSlipExpression2}b we understand that this cannot be. For distinct $f$, $\varphi$ can only meaningfully assume values within a $2\pi$-periodic interval (e.g., $\varphi\in[-\pi,\pi)$). If we instead want to decompactify $\varphi$, we may do so (as in Fig.~\ref{fig:PhaseSlipExpression2}a), but this cannot possibly create any new states. Here, we may represent the low-energy states by using the extended phase variable alone, without the need for an additional index $f$. This is the overcounting of states -- the root of the problems outlined above.

Let us limit the subsequent discussion to the compact description of $\varphi$ (Fig.~\ref{fig:PhaseSlipExpression2}b), and only occasionally refer to the extended picture. This may seem overly restrictive; after all, we have just illustrated how both representations are to quite some degree equivalent. However, there are two reasons for this step. First of all, we note that the re- or decompactification of the phase profile only works so easily for a single QPS junction. As we will discuss in a moment, 
for two QPS junctions in series the decompactification takes some nontrivial extra steps. Second, even if we consider only one QPS junction, as soon as we decompactify, $2\pi$-jumps in $\varphi$ would indeed have to be considered as non-linear capacitor elements, which (as we already indicated) we prefer to avoid, in order to be able to work with regular node-flux quantization.

To continue, we fix the Hamiltonian, such that it does not overcount the available low-energy states. In the compactified picture, the reduction of the Hilbert space can be readily performed by endowing the basis $\vert f\rangle$ with a phase dependence $\vert f\rangle\rightarrow \vert f\rangle_{\varphi}$, such that 
\begin{equation}\label{eq_boundary_f}
    \vert f\rangle_{\varphi\pm 2\pi}=\vert f\pm 1\rangle_\varphi\ .
\end{equation}
As a matter of fact, this $\varphi$-dependence of the phase slip states could have been derived also on the level of the minimization problem of the wire phase profile. Namely, if we understand $\vert f\rangle$ as a representation of the wave function of the phase profile inside the wire, then it becomes clear that the wave function must be $\varphi$-dependent in exactly the way we just stated (see also \SM{}, where we discuss this fact for a discrete realization of the QPS junction via Josephson junction arrays). If we now want to cast the physics related to $f$ into the framework of canonically conjugate variables, we have to explicitly keep the $\varphi$-dependence in the notation (at least initially -- see in a moment),
\begin{equation}\label{eq_H_new_QPS_alone}
    H_\text{QPS}^\text{c}=-E_S \cos\Big(\widehat{S}_\varphi\Big)+E_L\Big(\varphi+2\pi \widehat{f}_\varphi\Big)^2\ .
\end{equation}
Note that $e^{i \widehat{S}}=\sum_f \vert f\rangle \langle f-1\vert$ and $\widehat{f}=\sum_f f \vert f\rangle \langle f\vert$ still fulfill the ordinary commutation relations $[e^{i\widehat{S}},\widehat{f}]=-e^{i\widehat{S}}$, in spite of the $\varphi$-dependent basis.

\begin{figure}[htbp]
    \begin{center}
        \includegraphics[width=0.8\textwidth]{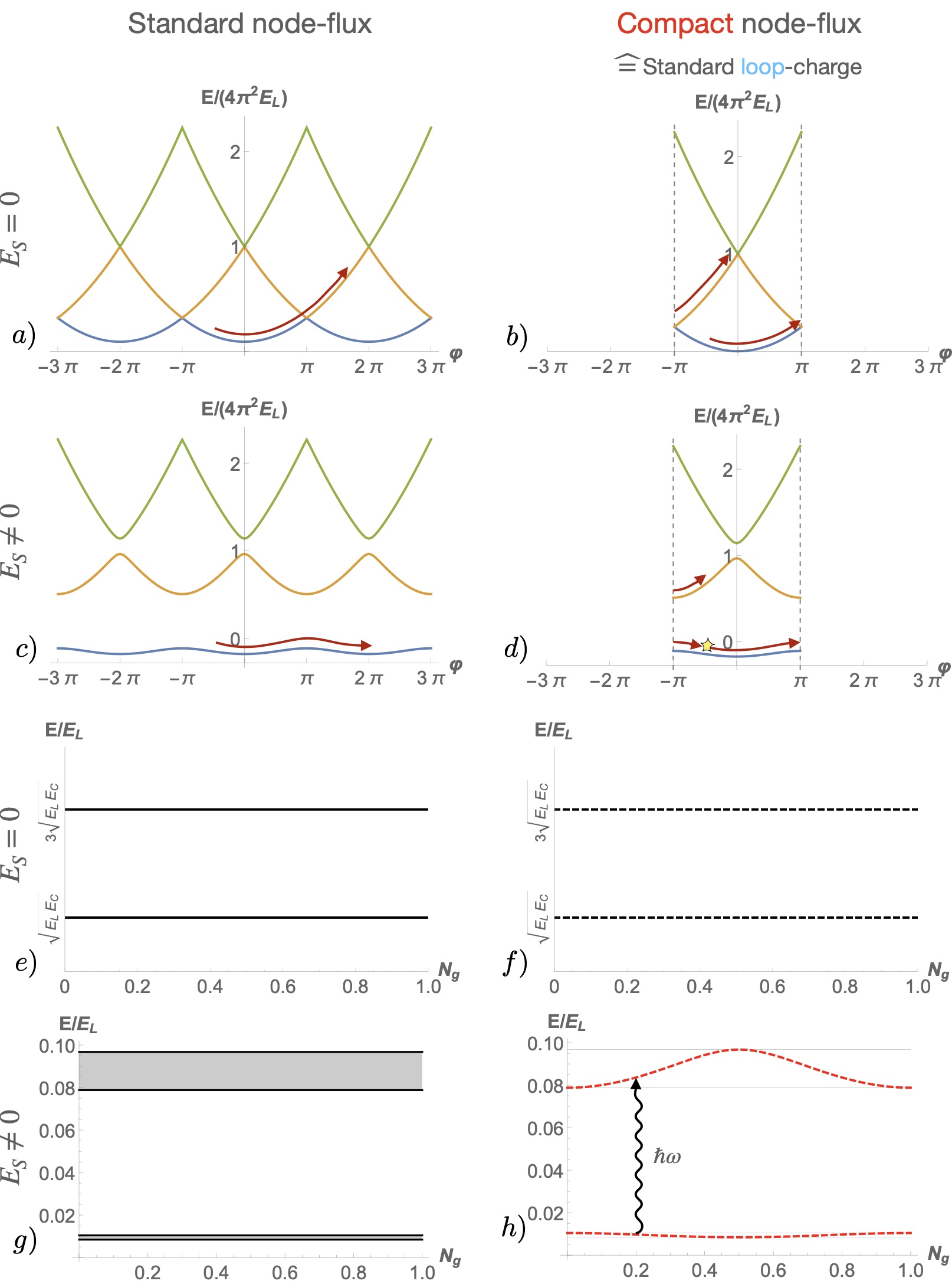}
    \end{center}
    \caption{Comparison of standard versus compact node-flux descriptions of the physics of QPS junctions. The energy spectrum of the standard node-flux treatment, Eq.~\eqref{eq_H_standard_QPS_alone}, is shown either for $E_S=0$ (a) or $E_S\neq 0$ (c). In (b \& d) we depict the spectrum of our newly proposed compact Hamiltonian, Eq.~\eqref{eq_H_new_QPS_alone}, for the same set of parameters. Locally, both Hamiltonians provide the same spectrum. Globally however, the standard picture yields periodically repeated copies of the same spectrum with extended $\varphi$. When shunting the QPS junction with a linear capacitance, see Fig.~\ref{fig:QPS_standard_treatment_start}b, the phase becomes dynamical, leading to various possible trajectories, represented by the red arrows. For $E_S=0$ [panels (a) and (b)] the dynamics of the standard and compact description are equivalent. In the standard description, the system simply stays within one parabola, belonging to a given phase slip state. Within the compact description, when the system state crosses, e.g., at  $\varphi=\pi$ it will transition to the next $f$-level, thus ultimately exploring the same parabola. Matters are different for finite $E_S$ [panels (c) and (d)]. Here, the compact description predicts trajectories where the system state can self-interfere (marked with a yellow star). Such trajectories do not exist in the standard description. As a consequence, the offset charge $N_g$ can no longer be removed by a gauge transformation (see main text). For a single QPS junction with a capacitive coupling, the physical picture of the compact node-flux formalism and the standard loop-charge formalism are the same: the latter follows from the former by means of a decompactification of the $\varphi$-space.
    }
	\label{fig:EnergiesWandWOQPS}
\end{figure}

Already at this stage, we are able to diagonalize $H_\text{QPS}$ and $H_\text{QPS}^\text{c}$ for a given (constant) $\varphi$. Crucially, both Hamiltonians provide us with exactly the same eigenspectrum for a given $\varphi$. This is to be expected, as locally, the $\varphi$-dependence can be regarded as a simple basis choice. However, there is a distinctly topological component, which cannot simply be removed by a unitary transformation. For $H_\text{QPS}$, we get infinite copies of the spectrum in the space of an unrestricted $\varphi$, whereas for $H_\text{QPS}^\text{c}$, the spectrum only lives strictly inside a $2\pi$-periodic interval, see Fig.~\ref{fig:EnergiesWandWOQPS}a,b,e,f. As such we have now established the necessary constraint on the available Hilbert space.

But while our approach successfully reduced the Hilbert space, it came at a cost of a significant complication: the explicit $\varphi$-dependence of the $f$-basis. How to correctly quantize with respect to $\varphi$, when incorporating the QPS Hamiltonian into a larger circuit?
Luckily, there is an argument why the local phase dependence can be transformed away. Namely, within the subspace of low-energy states $\vert f\rangle_\varphi$, the local $\varphi$-dependence can be neglected if the non-abelian Berry connection vanishes, $\langle f\vert\partial_\varphi \vert f'\rangle=0$ (this is a variation of a similar argument provided in Ref.~\cite{Kenawy_2022}, see also \SM{}). For the thin wire under consideration here, this seems to be indeed true to a very good approximation -- in fact, if we take the Nambu-Goldstone action as the field theory describing the wire, this Berry connection would as a matter of fact vanish exactly. In short, acting on one of the low-energy states $\vert f\rangle$ with the displacement operator $\partial_\varphi$ transports the state outside of the low-energy sector, into states which have been discarded by virtue of large $\omega_p$. This allows us to argue that it is legitimate, to locally ``flatten'' the low-energy basis, i.e., to assume $\vert f \rangle$ to be constant in $\varphi$. However, we still need to satisfy the global property, Eq.~\eqref{eq_boundary_f}. We do so by defining an arbitrary interval with length $2\pi$ for $\varphi$, e.g., $\varphi\in(-\pi,\pi]$, at the boundaries of which states with neighbouring $f$ are stitched together, according to Eq.~\eqref{eq_boundary_f}. Note however, that this step is an approximation, and will be studied in more depth in future dedicated works (see also final discussion).

We are thus finally at the point where we can combine all ingredients. Given a general circuit Lagrangian $L=T-V$~\cite{Devoret_1997,Burkard_2004,Vool_2017}, the full QPS Hamiltonian, Eq.~\eqref{eq_H_new_QPS_alone}, is added to the potential energy $V$. Within the potential energy, the local $\varphi$-dependence of the operators $\widehat{S}$ and $\widehat{f}$ shall be neglected (in accordance with our above argument of a vanishing Berry connection), such that the resulting Hamiltonian will look equivalent to the standard node-flux approach discussed above (simply by replacing $2\pi \wN_{S,j}\rightarrow \widehat{S}_j$ and $\wphi_{S,j}\rightarrow 2\pi \widehat{f}_j$). The crucial global constraint which was missing in the standard treatment, Eq.~\eqref{eq_boundary_f}, must then be included as an additional boundary condition on the wave function. The explicit form of the resulting boundary condition depends on the circuit model under consideration. First, let us spell it out explicitly for the device shown in Fig~\ref{fig:QPS_standard_treatment_start}b, with a single QPS junction. We denote the wave function for the QPS degrees of freedom $\varphi\in (-\pi,\pi]$ and $f\in\mathbb{Z}$ as $\psi_f(\varphi)$. The compact property of the $f$-basis, Eq.~\eqref{eq_boundary_f}, can then be accounted for by imposing
\begin{equation}\label{eq_boundary_psi}
    \psi_f(\pm\pi)=\psi_{f\pm1}(\mp\pi)\ .
\end{equation}
For instance, for the circuit given in Fig.~\ref{fig:QPS_standard_treatment_start}b, we end up with the Hamiltonian given in Eq.~\eqref{eq_H1QPS_node}, but importantly, including the constraint of Eq.~\eqref{eq_boundary_psi}. This constraint eliminates the Bloch wave vector, and restores the $N_g$-dependence of the energy spectrum, see Fig.~\ref{fig:EnergiesWandWOQPS}g and h. As indicated in the figure, without the constraint the theory gives rise to energy bands. A generic ac drive could therefore provoke both inter and intraband transitions. The constraint removes the bands and restores discrete levels, radically changing the ac response. The discrete bands and the $N_g$ dependence could have been obtained in the standard loop-charge formalism without constraints. Here, we are able to show easily how the two are related. Namely, the loop-charge picture can be obtained from our compact node-flux picture, simply by yet again unfolding from a compact $\varphi$ picture to the extended one, essentially by defining a new extended phase from the sum $\varphi+2\pi f\rightarrow \varphi$, see Fig.~\ref{fig:PhaseSlipExpression2}. If done correctly, this does not create spurious extra states in the Hilbert space, and the two formalisms are equivalent here.

Importantly, our formalism allows to formulate a precise relationship between the constraint on the Hilbert space and a generalized notion of charge quantization. This in turn provides an alternative, topological interpretation of the Aharonov-Casher effect, and refines the understanding of how qubit states in QPS-based circuits are read out and addressed (see at the end of this section). For this purpose, note that when rederiving the QPS Hamiltonian, we initially defined the total phase $\varphi$ as the phase inside the island. Since we have thus sharply defined the bounds of the island, the corresponding charge sitting on it (denoted in Fig.~\ref{fig:PhaseSlipExpression2}c as $N_q$), defined as the conjugate of $\varphi$, must have interger quantized eigenvalues as argued already in Ref.~\cite{Roman2021} for superconducting circuits, and similarly in Refs.~\cite{Aristov_1998,Gutman2010,Ivanov_2013,Riwar_2019b} for 0D and 1D electron systems. This can only be guaranteed by having $\varphi$ compact. But there may be other types of charge measurements, where a given charge detector fails to measure the island charge to such a precision that the quantization is visible. And in fact, the low-energy approximation of the QPS physics provides us naturally such a charge: when assuming the intrinsic dynamics of the QPS wire to be infinitely fast (large $\omega_p$), the phase profile inside the wire instantaneously follows $\varphi$, with exactly the linear dependence shown in Fig.~\ref{fig:PhaseSlipExpression2}a and b. That is, for low energies, displacing the phase on the island (the charge operator is nothing but a displacement operator in phase space) displaces with negligible delay the phase linearly inside the wire. That is, the relevant charge for the low energy dynamics is not the (quantized) charge $\wN_q$ inside the island, but a charge $\wN$ defined with a support reaching linearly into the wire, as shown in Fig.~\ref{fig:PhaseSlipExpression2}c (see also \SM{}). Since this charge has no sharp boundaries, it can (according to Ref.~\cite{Roman2021}) assume noninteger eigenvalues. Note similarly, that for the capacitive coupling to the island (see Fig.~\ref{fig:QPS_standard_treatment_start}b) we have not explicitly specified, with what spatial profile the gate actually couples to the island. For the low-energy regime here, this simply does not matter. Once we integrate out the dynamics within the wire, the charge $N$ is the only charge left, to which the capacitor could possibly couple. As a matter of fact, within the reduced low-energy Hilbert space we cannot even formulate a well-defined charge measurement of the original, quantized island charge $\wN_q$ anymore. Applying the quantized charge operator $\wN_q$ to the wave function transports it outside the low-energy sector of the Hilbert space in which it is defined. This is exactly the same feature which allowed us to locally flatten the $f$-basis and simplify the QPS dynamics to make the connection with the standard node-flux and loop-charge treatments. Ironically though, while the original quantized charge can no longer be defined within the low-energy theory, its presence is still felt in the form of the topological constraint on the wave function, which defines the dynamics of the circuit. This is an elaborate example of the statement formulated in Ref.~\cite{Roman2021}: while charge may not always be measured (or in this case, defined even) such as to see the quantization, it may nonetheless be indispensable to have a Hilbert space compatible with charge quantization to correctly predict the dynamics.

Let us now discuss the alternative interpretation of the Aharonov-Casher effect. First we note that within the loop-charge picture given in Eq.~\eqref{eq_H1QPS_loop}, we can understand the $N_g$-dependence of the energy spectrum as an Aharonov-Casher effect insofar as there is a gate-induced interference between a linear and a nonlinear capacitance. In our compactified formalism we get a different but equivalent picture. The state of the system returns to its original $\varphi$ position when going along the Brillouin zone, see Fig.~\ref{fig:EnergiesWandWOQPS}b and d, and can thus self-interfere -- picking up the phase $e^{i2\pi N_g}$ on the way. Importantly though, in order to actually be able to interfere, we have to have a finite $E_S$. Otherwise the system will just move along a single parabola, see Fig.~\ref{fig:EnergiesWandWOQPS}b. Hence, for $E_S=0$, there is no $N_g$-dependence in spite of the compactification, see Fig.~\ref{fig:EnergiesWandWOQPS}f. But for finite $E_S$, we find that the mechanism for the $N_g$-dependence is in fact the same as for the regular Josephson effect, see discussion after Eq.~\eqref{eq_H_Josephson_regular}. 

While for a single QPS junction, we can easily switch between a compact $\varphi$ and the state $f$ to an unbounded $\varphi$ and make contact with the standard loop-charge approach, we have to discuss what happens when including two QPS junctions in series, see Fig.~\ref{fig:QPS_junctions_standard_description_AB_AC}. Here our formalism returns the Hamiltonian,
\begin{equation}\label{eq_H_2QPS_compact}
    \begin{split}
    H_{C,\text{2QPS}}^\text{c}=E_C(\wN+N_g)^2-E_{S1}\cos(\widehat{S}_1)+E_{L1}(\wphi+2\pi\widehat{f}_{1})^2\\-E_{S2}\cos(\widehat{S}_2)+E_{L2}(\wphi+2\pi\widehat{f}_{2}-\phi_\text{ext})^2\ .
\end{split}
\end{equation}
Without any further ingredient, this would again map to the standard circuit node-flux description given in Eq.~\eqref{eq_H_2QPS_node}. However, in accordance with the principles we developed above, we have yet again to include a periodicity constraint. For the two coupled QPS junctions, the wave function $\psi_{f_1,f_2}(\varphi)$ now needs to satisfy
\begin{equation}\label{eq_psi_boundary_2QPS}
    \psi_{f_1,f_2}(\pm \pi)=\psi_{f_1\pm 1,f_2\pm 1}(\mp \pi)\ ,
\end{equation}
which is the generalization of Eq.~\eqref{eq_boundary_psi} for two QPS junctions in series. This additional constraint automatically removes the Bloch wave vector that we obtained for Eq.~\eqref{eq_H_2QPS_node}. Within the low-energy description of the QPS physics, the constraint can be readily understood: when we move $\varphi$ across the Brillouin zone, it creates phase slips (kinks or windings) simultaneously in both junctions.
Importantly, we can now with equal ease map the above Hamiltonian to the one obtained by means of the loop-charge formalism, and identify the missing flux quantization constraint. To that end, take Eq.~\eqref{eq_H_2QPS_compact} and perform the transformation,
\begin{align}
    \widehat{f}&=\frac{\widehat{f}_{1}+\widehat{f}_{2}}{2} &
\widehat{S}&=\widehat{S}_{1}+\widehat{S}_{2}\\
\delta f&=f_{1}-f_{2} &
\delta\widehat{S}&=\frac{\widehat{S}_{1}-\widehat{S}_{2}}{2}\ .
\end{align}
We arrive at
\begin{equation}
\begin{split}
    H_{C,\text{2QPS}}^{c}=E_{C}\left(\widehat{N}+N_{g}\right)^{2}-E_{S1}\cos\left(\widehat{S}/2+\delta\widehat{S}\right)+E_{L1}\left(\widehat{\varphi}+2\pi\widehat{f}+\pi\delta\widehat{f}\right)^{2}\\-E_{S2}\cos\left(\widehat{S}/2-\delta\widehat{S}\right)+E_{L2}\left(\widehat{\varphi}+2\pi\widehat{f}-\pi\delta\widehat{f}-\phi_{\text{ext}}\right)^{2}.
\end{split}
\end{equation}
We can now safely decompactify: as just noted, due to the nontrivial boundary conditions, the progression with the capacitive element across the boundaries of the Brillouin zone provokes a simultaneous slip of $f_1$ and $f_2$ by $\pm 1$. In terms of the new variables, this would be a change of $f$ by $\pm 1$, whereas $\delta f$ would stay the same. We can thus combine the compact $\varphi\in[\pi,\pi)$ with the $f$ index to an extended $\varphi$, and replace the operator $\widehat{S}$ with $2\pi \wN$. Hence, we arrive at a Hamiltonian equivalent to Eq.~\eqref{eq_H_2QPS_loop} within the loop-charge approach. To see this, perform in Eq.~\eqref{eq_H_2QPS_loop} the coordinate transformations $N^\circ=N_1^\circ-N_2^\circ$ and $\delta N^\circ=(N_1^\circ+N_2^\circ)/2$,  and equate $2\pi\delta N$ to $\delta S$. Except that with our treatment, we directly arrive at the correct quantization of the flux, since $\delta f$ is defined as an integer right from the get-go. Note that this is also a computational advantage for numerical evaluations of the eigenspectra of circuits, since we can work with a significantly smaller Hilbert space.

\begin{figure}
    \centering
    \begin{center}
        \includegraphics[width=1.0\textwidth
        ]{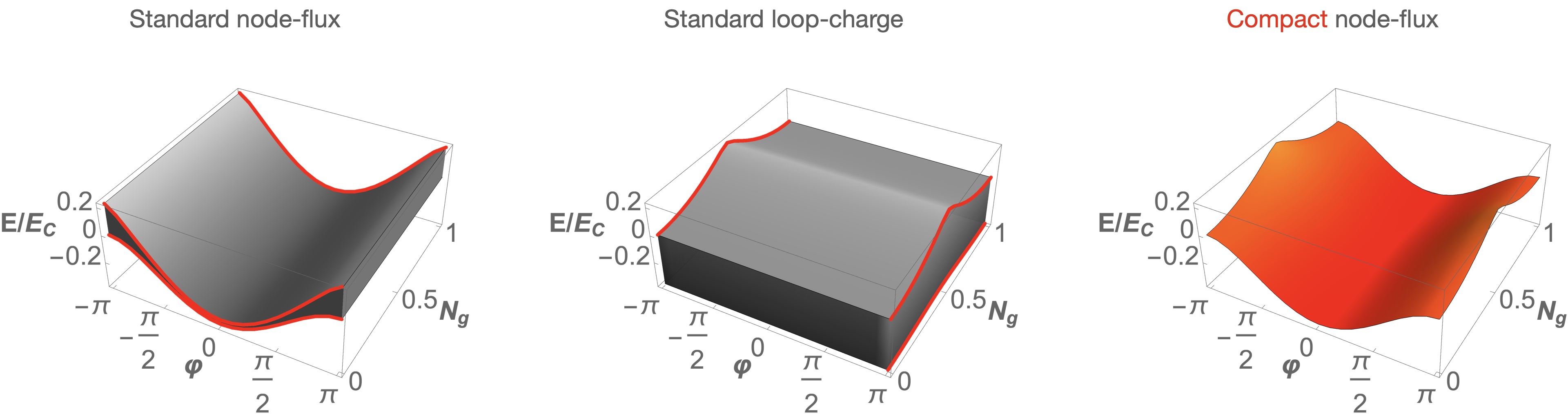}
    \end{center}
    \caption{Different spectra predicted for a circuit with a central island, coupled to two QPS junctions and a central gate, see Fig.~\ref{fig:QPS_junctions_standard_description_AB_AC}, subject to a gate-induced offset charge $N_g$ and an externally applied flux $\phi_\text{ext}$. (a) The standard node-flux approach predicts a spectrum with continuous bands, depending explicitly on $\phi_\text{ext}$ but not on $N_g$. (b) The reverse is true for the loop-charge formalism. (c) Our compact node-flux approach removes the bands and predicts discrete energy levels, due to the topological constraint in the Hilbert space, relying on charge quantization. The spectrum depends on both $N_g$ and $\phi_\text{ext}$, and thus unifies the Aharonov-Bohm and Aharonov-Casher effects. }
	\label{fig:standartVSloopVSCompact}
\end{figure}

To summarize, we find that our treatment fixes the Bloch wave vectors of both the standard node-flux and the standard loop-charge formalisms. The resulting differences in the spectra is shown in Fig.~\ref{fig:standartVSloopVSCompact} (for the calculation, see \SM{}). By means of the compactification of the Hilbert space, instead of the respective Bloch bands we recover a discrete energy spectrum, depending on both the gate-induced offset charge $N_g$ and the externally applied flux $\phi_\text{ext}$, unifying the Aharonov-Bohm and Aharonov-Casher effects. As for the latter, we finally see that within our formalism, its origin is no different from the circuit with a single QPS junction. In the picture of a compactified $\varphi$, the system picks up the phase $e^{i2\pi N_g}$ when covering the Brillouin zone once, in analogy to the process shown in Fig.~\ref{fig:EnergiesWandWOQPS}d (except that now, there will be two indices $f_1,f_2$ instead of the single $f$).

While this alternative picture for the Aharonov-Casher effect is equivalent to the one promoted by the standard loop-charge approach (where for two QPS junctions, we have essentially an Aharonov-Casher type interference between 3 capacitors, two nonlinear and one linear, see Eq.~\eqref{eq_H_2QPS_loop}), it may nonetheless offer some formal advantages. Note that the Hamiltonian most often deployed for a circuit with two QPS junctions, is not the full Hamiltonian with the extra capacitive energy term, but the simplified Hamiltonian in Eq.~\eqref{eq_H_2QPS_loop_AC}, which, as we pointed out above, usually does not come with a well-formulated derivation. Hence, especially Eq.~\eqref{eq_H_2QPS_loop_AC} gives the wrong impression that one would need a minimum of two QPS junctions to see Aharonov-Casher interference patterns (just as one would need two Josephson junctions for a SQUID). Our formalism, however, puts the observed $N_g$-dependence on the exact same level as for regular Josephson junctions. To emphasize this, notice also the simplification of the QPS junction for $E_S\gtrsim E_L$ to an effective regular Josephson effect, Eq.~\eqref{eq_EJeff}. Due to the compactness of $\varphi$, the $N_g$-oscillations of the low energy spectrum are in this regime very explicitly the same as for a regular charge qubit~\cite{Cottet_2002}, see also \SM{}.

\subsection{Reduced computational space due to dual frustration}

One final aspect that we want to touch upon, is that determining the correct Hilbert space size is not only crucial for predicting the right spectrum and circuit dynamics, but also to correctly assess the computational space when examining the utility of circuits for potential qubit realizations. In particular, we pick up on an interesting recent proposal~\cite{Thanh2019} to use QPS junctions for the realization of the Gottesman-Kitaev-Preskill (GKP) code~\cite{Gottesman_2001}. The proposed device replaces one of the two QPS junctions in the circuit shown in Fig.~\ref{fig:QPS_junctions_standard_description_AB_AC} by a regular Josephson junction.
The authors of Ref.~\cite{Thanh2019} derive the following Hamiltonian~\footnote{This is one of the few cases, where the non-linear capacitor treatment within a node-flux approach works (see above mentioned issues about non-convexity of the kinetic energy term), and the authors find a Hamiltonian equivalent to our standard node-flux approach.}
\begin{equation}
    \label{eq_H_standard}
    \begin{aligned}
    H_{C,\text{QPS},J}=&-E_{S}\cos\left(2\pi\widehat{N}_{S}\right)+E_{L}\left(\widehat{\varphi}_{S}-\widehat{\varphi}\right)^{2}\\&+E_{C}\left(\widehat{N}+N_{g}\right)^{2}-E_{J}\cos\left(\widehat{\varphi}-\varphi_{e}\right)\ ,
    \end{aligned}
\end{equation}
In our compact node-flux approach, we again replace the operators $2\pi \wN_S\rightarrow \widehat{S}$ and $\wphi_S\rightarrow 2\pi \widehat{f}$ (to emphasize that $f$ is integer), and in addition, impose the symmetry constraint in $(\varphi,f)$-space on the wave function for one QPS junction, given in Eq.~\eqref{eq_boundary_psi}.

Yet again, the seemingly harmless compactification of $\varphi$ has profound consequences. One of the most interesting regimes in Ref.~\cite{Thanh2019} is for large $E_L$. As already foreshadowed, this limit does not work as the authors of~\cite{Thanh2019} expect. Without any constraints, one might mistake the $E_L\rightarrow \infty$ limit to serve as a Lagrange multiplier, perfectly coupling $\varphi_S=\varphi$ and thus promoting the auxiliary charge $N_S$ to the physical charge $N$. This would indeed realize the double-cosine Hamiltonian required to implement important features of the GKP code. Our compact formalism shows that there is simply not enough available Hilbert space to reach this limit. The effect can be thought of as a dual frustration: in reality $f$ is integer, and $\varphi$ is restricted to an interval of size $2\pi$. Hence, the two cannot be coupled by rendering $E_L$ large, because they are fundamentally incompatible. Instead, for $E_L\gg E_S$, QPS are simply suppressed, and the circuit begins to work like a regular fluxonium~\cite{Koch_2009}.

Consequently, if one wants to preserve any hope of realizing a double-cosine Hamiltonian one has to search in the opposite regime, where $E_L$ is small compared to $E_S$. We expect that this first of all a significant experimental hurdle, as $E_S$ cannot always tuned to large values for all physical implementations of QPS wires. In Ref.~\cite{Houzet_2019}, it is predicted that large $E_S$ may be reached in very large Josephson junction arrays due to disorder in the local offset charge profile. Note furthermore, that for $E_L,E_C\rightarrow 0$, strictly speaking, the operators inside the two cosines are not even the actual canonically conjugate pair of operators. The operator $\widehat{S}$ knows about the phase slip state inside the wire, and $\varphi$ represents the phase inside the island. This problem in and of itself can, at least in principle, be solved by the above introduced decompactification procedure, see Fig.~\ref{fig:PhaseSlipExpression2}. We thus eliminate again the $S,f$ degrees of freedom and end up with a single pair of $N,\varphi$ where $\varphi$ is now no longer compact. This seems to salvages the idea of realising a GKP code. But note how the new (non-compact) phase and charge variables are defined: they extend linearly into the wire, see Fig.~\ref{fig:PhaseSlipExpression2}c. Now remember, that the central idea behind the GKP code as a quantum error correction code is to track random external shifts (here in $N_g$ and $\phi_{ext}$) by appropriate nonlocal measurements in the respective spaces of the conjugate observables, without destroying the quantum information. That is, merely looking at the Hamiltonian lures us into the false impression that the analysis of the qubit state can be performed by measurements of charge and phase on the island. The physical picture we develop demonstrates to the contrary, that we instead need to measure the state inside the QPS wire. Thus, the above realization is another important caveat, significantly complicating the idea of an experimental implementation of quantum error correction strategies by using QPS junctions.

\section{Discussion} 

We demonstrate that it is possible to regard QPS junctions as purely inductive elements, thus rendering them compatible with regular node-flux quantization. We derive an important constraint on the wave function, based on charge quantization, which explicitly reduces the Hilbert space, and eliminates spurious degrees of freedom. As we show, this has profound consequences on how the circuit can interact with externally applied electric and magnetic fields, and allows us to unify the Aharonov-Bohm and Aharonov-Casher effects within one formalism. Furthermore, the reduction of the Hilbert space is an important principle when examining the utility and feasibility of possible qubit architectures involving QPS junctions.

We note that the constraint can be regarded as a minute, but decisive breaking of the exact duality between the Josephson effect and quantum phase slips. This is due to the fundamental difference with respect to how charge and phase degree of freedoms enter the Hamiltonian (in particular, the latter enters always in the exponent, $e^{i\phi}$). To predict the correct physics, it appears crucial that the theory retains a possibility for the phase to be represented on a compact manifold, even when charge quantization is not explicitly present (or removed within the low-energy sector of the state space).

Last but not least, our treatment of QPS physics will give rise to a number of new research directions. On the one hand, the local flattening of the $f$ basis (i.e., removing the $\varphi$-dependence) is only approximately true. This approximation will be scrutinized in more detail, and with more sophisticated methods in the future. Note in particular the predicted dependence of $E_S$ on local offset charges in Josephson junction arrays~\cite{Pop2012,Houzet_2019}. For such array models, the unitary can be understood as an additional, dynamical shift of the offset charge, which we expect to give rise to a $\dot{\varphi}$-dependence of $E_S$, massively increasing the complexity of the overall circuit dynamics. Moreover, it will be interesting to bring topological superconductors into the mix, where the presence of Majorana bound states could potentially change the width of the Brillouin zone, i.e., $\varphi$ may live on an interval of size $4\pi$ instead of $2\pi$. A dedicated work tackling that particular question is underway, where we expect to see competing effects of $2\pi$- versus $4\pi$-periodicity. Moreover, the physics of 1D superconducting structures (such as Josephson junction arrays) are often mapped onto the sine-Gordon model~\cite{Gurarie_2004,Houzet_2019}, where disorder may lead to interesting renormalization effects~\cite{Houzet_2019,Giamarchi_1988}. However, the sine-Gordon equation has extended quantum fields. We are so far unaware of theoretical studies examining the possibility of a modified sine-Gordon model with compact quantum fields -- which we believe to be of relevance to model superconducting structures, and (since charge quantization is a fundamental property) possibly even beyond.

\section*{Acknowledgements}

We warmly thank David DiVincenzo and Fabian Hassler for highly stimulating discussions. We are further indebted to Gianluigi Catelani, Ioan Pop and Alex Kashuba for additional discussions and inputs. This work has been funded by the German Federal Ministry of Education and Research within the funding program Photonic Research Germany under the contract number 13N14891.

\bibliography{bibliography,biblio}

\end{document}


\title{Supplementary Material: Compact description of quantum phase slip junctions}
\author{C. Koliofoti and R.-P. Riwar}
\affiliation{Peter Gr\"unberg Institute, Theoretical Nanoelectronics, D-52425 J\"ulich, Germany }

\maketitle

\section{Loop-charge quantization for circuits involving one and two QPS junctions}

In the main text, we give the loop-charge Hamiltonians of the two circuits of Fig.~\ref{fig:QPS_standard_treatment_start} and Fig. \ref{fig:QPS_junctions_standard_description_AB_AC}. {Here we briefly show how they can be derived.} In both cases we will {use} Fig.~3 of Ref.~\cite{Ulrich_2016} {to build the Lagrangian}.

The first circuit (Fig.\ref{fig:QPS_standard_treatment_start}) consists of a QPS junction, an inductor and a  gate voltage with a capacitance {in a loop}. The Lagrangian for this system {reads}
\begin{equation}
	\begin{split}
		 L^\circ=E_S \cos\lr{2\pi N^\circ}+\frac{L}{2} \lr{2e}^2 \dot{N}^{\circ}{}^{2}\\-\frac{\lr{2e}^2}{2C_g} N^{\circ}{}^{2} - 2e V_g  N^{\circ} \ .
	\end{split}
\end{equation}
First we {have to} find {the canonically conjugate momentum to the loop-charge $N^\circ$, }$\varphi^\circ$,
\begin{equation}
    \label{eq:phicircLagrangian}
    \varphi^\circ=\frac{\partial L^\circ}{\partial \dot{N}^{\circ}}= L \lr{2e}^2 \dot{N}^{\circ}
\end{equation}
Thus we can now find the Hamiltonian {by means of the Legendre transformation}
\begin{equation}
    \begin{aligned}
    H^{\circ}=&\dot{N}^{\circ}\varphi^{\circ}-L^{\circ}\\
    =&\frac{1}{2L}\left(\frac{\varphi^{\circ}}{2e}\right)^{2}-E_{S}\cos\left(2\pi N^{\circ}\right)-\frac{C_{g}}{2}V_{g}^{2}\\ & \qquad+\frac{\left(2e\right)^{2}}{2C_g}\left(N^{\circ2}+\frac{2C_{g}}{\left(2e\right)}V_{g}N^{\circ}+\left(\frac{C_{g}}{\left(2e\right)}V_{g}\right)^{2}\right)\ .
    \end{aligned}
\end{equation}
By setting $E_L=1/\lr{4e^2L}$, $E_C={\left(2e\right)^{2}}/{2C_g}$ and $N_g = {C_g V_g}/{\left(2e\right)}$  this leads us exactly to the Hamiltonian of Eq. \eqref{eq_H1QPS_loop}.

The same calculation can be done for the circuit in Fig. \ref{fig:QPS_junctions_standard_description_AB_AC}, consisting of two QPS junction, two inductor and a  gate voltage with a capacitance. Again we will follow the recipe of Ref.~\cite{Ulrich_2016} to construct the Lagrangian
\begin{equation}
    \begin{split}
    L^{\circ}=E_{S1}\cos\left(2\pi N_{1}^{\circ}\right)+E_{S2}\cos\left(2\pi N_{2}^{\circ}\right)+\frac{L_{1}}{2}\left(2e \dot{N}_{1}^{\circ}\right)^{2}\\+\frac{L_{2}}{2}\left(2e\dot{N}_{2}^{\circ}\right)^{2}-\frac{\left(2e\right)^{2}}{2C_g}\left(N_{1}^{\circ}-N_{2}^{\circ}\right)^{2}-2e V_{g}\left(N_{1}^{\circ}-N_{2}^{\circ}\right) \ .
    \end{split}
\end{equation}
{The canonically conjugate phase} $\varphi^\circ_i$ will be of the form of $\varphi^\circ$ in Eq.~\eqref{eq:phicircLagrangian} with $N^\circ\rightarrow N^\circ_i $ and $L\rightarrow L_i $. Finally we can construct the Hamiltonian {again through}
\begin{equation}
    \begin{aligned}
    H^{\circ}=&\dot{N}_{1}^{\circ}\varphi_{1}^{\circ}+\dot{N}_{2}^{\circ}\varphi_{2}^{\circ}-L^{\circ}\\
    =&\sum_{i\in\left\lbrace1,2\right\rbrace}\lrB{\frac{1}{L_{i}}\left(\frac{\varphi_{i}^{\circ}}{2e}\right)^2-E_{Si}\cos\left(2\pi N_{i}^{\circ}\right)}
    \\&\qquad+\frac{\left(2e\right)^{2}}{2C_g}\left(N_{1}^{\circ}-N_{2}^{\circ}+N_{g}\right)+\text{const.}
    \end{aligned}
\end{equation}
This leads us to the Hamiltonian of Eq.~\eqref{eq_H_2QPS_loop}.


\com{
\section{Calculation of eigenspectra and density of states}\label{sec_DOS}

In the main text, we have discussed the differences in the eigenspectrum when describing the phase slip junction either in its standard form $\sim\cos(2\pi \widehat{N})$ or via our proposed version $\sim\cos(\widehat{S})$. We here make the explicit calculations for the different eigenspectra and briefly explain how we obtained the corresponding density of states (DOS) in Figs.~\ref{fig:PhaseSlipJJMJ}c and d.

\subsection{Standard description of QPS junction}
Let us derive the eigenenergies of Hamiltonians of the form \eqref{eq_H_cos_cos} with $\widehat{X}\rightarrow \widehat{N}$ and $\widehat{P}\rightarrow \widehat{\varphi}$ (corresponding to the standard description of the QPS junction), for two values of $q/p$.
First we consider a shunt with a regular Josephson Junction (JJ), where $q/p=1$, with the Josephson energy $E_J=E_{J,1}$,
  \begin{equation}
    \label{eq:HamShuntJJ}
 	H= -E_S \cos\left(2\pi \wN \right) - E_J \cos\left(\wphi\right).
 \end{equation}
For this case the eigenenergies are given by Ref.~\cite{Zak1967} as
\begin{equation}
    \label{eq:EnShuntJJ}
	E= -E_S \cos\left(2\pi k\right) - E_J \cos\left(\varphi_0\right),
\end{equation}
with $\varphi_0 \in \left(-\pi,\pi\right]$ and $k\in\left(-1/2,1/2\right]$. Let us compute the DOS for the case $E_S=E_J$. Rescaling the energy as $\epsilon\equiv E/E_S$ as well as $x=2\pi k$ and $y=\varphi_0$, we define the normalized density of states as
\begin{equation}
    \rho(\epsilon)=\int_{-\pi}^\pi\frac{dx}{2\pi}\int_{-\pi}^\pi\frac{dy}{2\pi}\delta[\epsilon-\cos(x)-\cos(y)]\ .
\end{equation}
Performing the integrations successively, we arrive at the analytic result
\begin{equation}\label{eq_DOS_analytic}
    \rho(\epsilon)=\frac{4}{\pi^2}\frac{1}{|\epsilon|}\text{Re}\left[iK\left(\frac{4}{\epsilon^2}\right)\right]\ ,
\end{equation}
where $K(x)$ is (a variation of) the complete elliptic integral of the first kind, defined as $K(x)=\int_0^{\pi/2}d\theta \left[1-x \cos^2(\theta)\right]^{-1/2}$. On Mathematica, this function is defined as EllipticK.

A shunt with a Majorana junction on the other hand is described by the parameter $q/p=1/2$ and the energy $E_M=E_{J,1/2}$, leading to
 \begin{equation}
    \label{eq:HamlitonianShuntedMJ1}
    H= -E_S \cos\left(2\pi \wN \right) - E_M \cos\left(\frac{\wphi}{2}\right).
 \end{equation}
 To diagonalize, we use the ansatz for the wave function in $\varphi$-space
\begin{equation}
    \label{eq:AnsatzES1ShuntedMJ}
    \begin{aligned}
 	    \psi \left(\varphi\right) =&a\sum_{l\in\text{even}} e^{ik\varphi}\delta\left(\varphi-\varphi_0 - 2 \pi l \right)  \\
 	    &\quad+b\sum_{l\in\text{odd}} e^{ik\varphi}\delta\left(\varphi-\varphi_0 - 2 \pi l \right) \ .
    \end{aligned}
\end{equation}
When acting with the Hamiltonian on this wave function ansatz, we find that the eigenenergies $E$ need to fulfill
\begin{align*}
	a E&=- b E_S \cos\left(2\pi k\right) - a E_M \cos\left(\varphi_0/2\right)\\
	b E&=- a E_S \cos\left(2\pi k\right) + b E_M \cos\left(\varphi_0/2\right)
\end{align*}
Therefore, we obtain the eigenenergies:
\begin{equation}
    \label{eq:EEShuntedMJ1}
     E=\pm\sqrt{E_{S}^{2}\cos^{2}\left(2\pi k\right)+E_{M}^{2}\cos^2\left(\varphi_0/2\right)} \ .
\end{equation}
Let us again choose $E_S=E_M$ and rescale similarly $\epsilon=E/E_S$, $x=2\pi k$ as well as $y=\varphi_0/2$, to compute the normalized DOS
\begin{equation}
    \rho(\epsilon)=\int_{-\pi}^\pi\frac{dx}{2\pi}\int_{-\pi}^\pi\frac{dy}{2\pi}\delta[\epsilon\mp\sqrt{\cos^2(x)+\cos^2(y)}]\ .
\end{equation}
Similar integration steps as before now lead to
\begin{equation}\label{eq_DOS_analytic_MJ}
    \rho(\epsilon)=\theta(2-\epsilon^2)\frac{2}{\pi}\left|\frac{\epsilon}{\epsilon^2-1}\right|K\left[1-\frac{1}{\left(\epsilon^2-1\right)^2}\right]\ ,
\end{equation}
with the same complete elliptic integral $K$. This function is shown in Fig.~\ref{fig:PhaseSlipJJMJ}c in the main text.

\subsection{Alternative description of QPS junction}
In the main text, we present an alternative description for the QPS junction, by means of the canonically conjugate pair of operators $[\widehat{S},\widehat{f}]=i$. The resulting alternative Hamiltonian is
\begin{equation}
    H=-E_S\cos\left(\widehat{S}\right)-E_{J,q/p}\cos\left(\frac{q}{p}\widehat{\varphi}\right)\ ,
\end{equation}
which, when expressed in the eigenbasis of $\widehat{f}$, is given as
 \begin{equation}
 	H=-\frac{E_S}{2}\sum_{f}\left(\left|f+1\right\rangle \left\langle f\right|+\text{h.c.}\right)-E_{J,q/p}\cos\left(\frac{q}{p}\wphi\right) .
\end{equation}
Since the operator $\widehat{S}$ does not act on the $\varphi$-space, we can directly solve for all values of $q/p$, done by rescaling $\frac{q}{p}\varphi\rightarrow \widetilde{\varphi}$.
For the wavefunction we separate the $\varphi$-solution and the $f$-solution, thus using the ansatz
\begin{equation}
	\label{eq:AnsatzES2ShuntedJJ}
	\left|\psi\left(\varphi\right)\right\rangle =
	\delta\left(\widetilde{\varphi}-\widetilde{\varphi}_0\right)\otimes \sum_{f}e^{i2\pi kf}\left|f\right\rangle .
\end{equation}
This ansatz directly solves the stationary Schr\"odinger equation, resulting the same type of eigenspectrum as in \eqref{eq:EnShuntJJ},
\begin{equation}
	E= -E_S \cos\left( 2 \pi  k\right) - E_{J,q/p} \cos\left(\widetilde{\varphi}_0\right)\ ,
\end{equation}
except that now, the solution is valid for all values of $q/p$. When yet again considering $E_S=E_{J,q/p}$ and rescaling the energy and other parameters, we arrive at the same DOS as in Eq.~\eqref{eq_DOS_analytic}. Consequently, the DOS for $q/p=1/2$, corresponding to the shunt with the Majorana junction, provides a DOS distinct from Eq.~\eqref{eq_DOS_analytic_MJ}. This differing DOS is depicted in Fig.~\ref{fig:PhaseSlipJJMJ}d in the main text.

}

\section{Finite element model and differing charge definitions}

\begin{figure}[tbp]
    \begin{center}
        \includegraphics[width=0.45\textwidth]{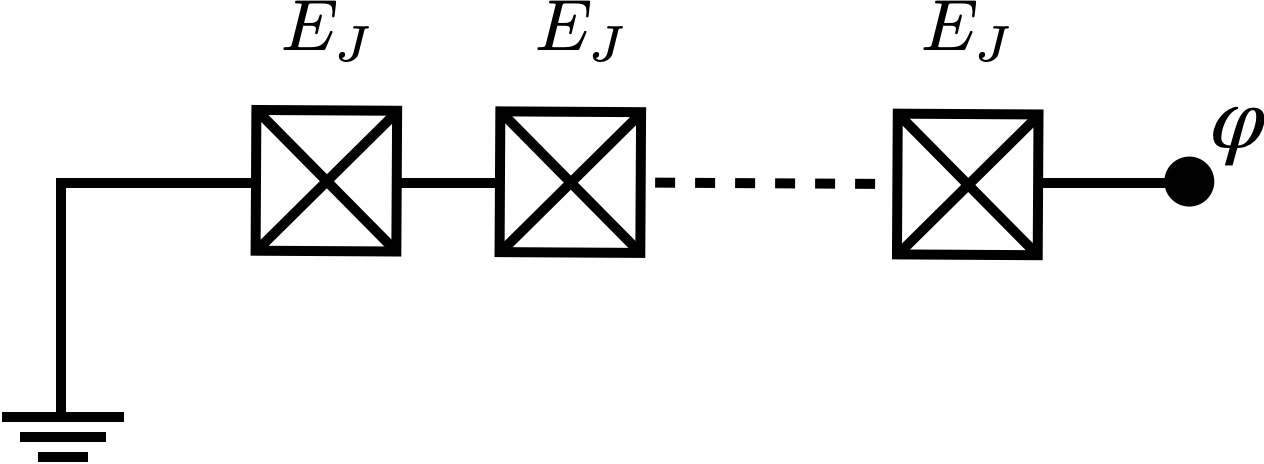}
    \end{center}
    \caption{Josephson junction array. Note that the individual islands may be in general connected by a highly complex capacitive network. For our considerations the details of the capacitive coupling is irrelevant, which is why we do not explicitly show it here. Also note that if we use the JJ array model to simulate a bulk superconducting wire, then the circuit dynamics will likely be dominated by the capacitances to ground. }
	\label{fig:JJarrayWithMajorana001}
\end{figure}

{In the main text, we revisit the low-energy physics of QPS junctions. In this section we illustrate some of the conclusion of the main text, by means of a finite element approach, when the wire is modelled by a discrete chain of Josephson junctions, see Fig.~\ref{fig:JJarrayWithMajorana001}. We use this model also to explain how different basis choices for the low-energy states of the QPS junction correspond to the measurement or interaction with different charges $N$ and $N_q$ as hinted at in the main text.}

For this purpose, we decompose the QPS wire into a finite but large number islands indexed by $j=1,\ldots,J$, hosting charges and phases $[\widehat{N}_j,\widehat{\varphi}_j']=i\delta_{jj'}$. Here the last phase $\varphi:=\varphi_J$ is kept arbitrary but constant.   
The islands may be capacitively coupled (either among themselves or to ground) and exchange Cooper pairs formally via JJs. The ends of the resulting JJ array are going to be phase-biased {due to a finite $\varphi$ and choosing $\varphi_0=0$ (contact to ground)}.
The total Hamiltonian is given as $H=T+V$, where we cast the entirety of the capacitive energy into the kinetic energy term
\begin{equation}\label{eq_T_of_H}
    T=\frac{(2e)^2}{2}\sum_{j,k=1}\widehat{N}_{j}\left(C^{-1}\right)_{jk}\widehat{N}_{k}\ ,
\end{equation}
and all Josephson energies into a so-called potential energy term
\begin{equation}
    \label{eq:HamJJarrayshuntMJ}
    \begin{split}
        V=- E_J \sum_{j=1}^J \cos\left(\widehat{\varphi}_j-\widehat{\varphi}_{j-1}\right) \ .
    \end{split}
\end{equation}
{While this model is obviously precise if the QPS junction is actually made of a Josephson junction array, we believe that it should retain some validity in a continuous limit, i.e., for a thin wire.}

\subsection{Derive low-energy Hamiltonian in Eq.~\eqref{eq_H_new_QPS_alone}}

As for the influence of the capacitive contribution to the circuit dynamics, we notice the following. With all Josephson energies the same, and at sufficiently low-energies, the potential energy (the Josephson-energy contributions of the JJ array) have well-defined local minima. From this perspective, the capacitances merely provide a mechanism whereby the system can perform quantum jumps between different local minima. These are the quantum phase slips. For our purposes, we do not need to concern ourselves with the details of the capacitance matrix, so long as they provide this quantum tunneling mechanism. Ultimately the capacitive energy will provide a means to estimate the energy associated to jumps between different minima, which has already been done for JJ arrays, see Refs.~\cite{Matveev_2002,Gurarie_2004,Houzet_2019}.

Let us therefore evaluate the energy minima of the potential energy. For this purpose, we may consider all phase variables as classical parameters $\widehat{\varphi}_j\rightarrow \varphi_j$.
We can deploy the minimization scheme used in Refs.~\cite{Catelani_2011,Roman2021}.
For fixed $\varphi:=\varphi_J$, and large $J$, the minimization for the remaining $\varphi_j$ (with $j\neq 0, J$) results in
\begin{equation}
	\label{eq:MinAnsatzJJarray}
	\ensuremath{\varphi_{j=1\ldots J-1}^{\left(0,f\right)}=\frac{j}{J}\left(\varphi_{}+2\pi f\right)\mod2\pi}\ .
\end{equation}
In order to obtain the energy at the minima, we approximate the potential energy $V$ \eqref{eq:HamJJarrayshuntMJ} for small phase differences $\varphi_j-\varphi_{j-1}\ll 1$ as
\begin{equation}
    \begin{aligned}
        V&\approx\text{const.}+ \frac{E_J}{2} \sum_{j=0}^J \left(\varphi_j-\varphi_{j-1}\right)^2 \ .
    \end{aligned} 
\end{equation}
Importantly, due to the potential energy being periodic in all phases, the minima can be defined up to modulo $2\pi$. This corresponds to the compact representation of the local minima, see Fig.~\ref{fig:PhaseSlipExpression2}b in the main text. Plugging in \eqref{eq:MinAnsatzJJarray} into the approximated potential energy will yield
\begin{equation}
     \begin{aligned}
        V&\approx 
        \frac{E_{J}}{2}\frac{1}{J}\left(\varphi+2\pi f\right)^{2}
    \end{aligned} 
\end{equation}
The first quadratic term corresponds to the inductive energy appearing in the main text in Eq.~\eqref{eq_H_new_QPS_alone}. 
The Josephson junction array has an effective array inductance $E_L\equiv E_J/J$. In the continuous limit we find $E_L\rightarrow 1/(4e^2 l L)$, with the array (wire) length $L=\Delta x J$, where $\Delta x$ is the distance between two junctions.
The quantum phase slips between different minima $f$ are in the main text described by the operators $e^{\pm i\widehat{S}}\equiv \sum_f |f\pm 1\rangle_{\varphi_J}\langle f|_{\varphi_J}$. The feature usually neglected is the $\varphi$-dependence of the $f$-basis. Namely, we can think of the minima solutions as wave functions centered around the minima. This gives rise to the $\varphi$-dependent QPS operators.

\subsection{Elimination of local Berry connection}

{In the main text, we argued that locally, we can neglect the phase dependence, if the Berry connection term $-iU\partial_\varphi U^\dagger$ vanishes. This is approximately true when the Josephson energies are sufficiently large with respect to the capacitive energies. Then, the wave functions of the local minima resemble the local ground states of the harmonic oscillator chain. Applying the derivative $\partial_\varphi$ maps the local ground states to excited states, such that in the low-energy sector, where we keep only the local ground states, we indeed find $-iU\partial_\varphi U^\dagger\approx 0$. This is however only approximately true. We see that if we included the full Hilbert space with the excited states, the $\varphi$-dependence of the basis will start to play a role. This will be examined in more detail in future work.
}

\com{
Finally, we extend the minimization procedure such that $\varphi_J$ is variable, too. This is merely to verify that the above minimization procedure works even in the presence of the Majorana shunt, i.e., that the correction due to finite $E_M$ is small. At the local minima the derivative $\partial_{\varphi_J}$ should vanish, thus we obtain the condition
\begin{equation}
    \label{eq:ForMinimazation001}
    E_{J}\frac{1}{J}\left(\varphi_{J}+2\pi f\right)+\frac{E_{M}}{2}\sin\left(\frac{\varphi_{J}-\varphi_{\text{ext}}}{2}\right) =0
\end{equation}
For large $J$, such that $E_J /J \ll E_M$, we minimize the $\sin$-term, with the ansatz $\varphi_J=\varphi_\text{ext}+\delta\varphi_J$, with $\delta\varphi_J\ll 1$.
Inserting this in \eqref{eq:ForMinimazation001}, we get the correction as a function of $f$ as,
\begin{equation}
    \begin{aligned}
        \delta\varphi_{J}(f)\approx-\frac{4E_{J}}{JE_{M}}\left[\varphi_{\text{ext}}+2\pi f\right]
    \end{aligned}
\end{equation}
Therefore, we obtain a more general minimization for the potential term of the Hamiltonian \eqref{eq:HamJJarrayshuntMJ} including a finite $E_M$, that reads:
\begin{equation}
	\label{eq:MinimaJJarray001}
	\begin{aligned}
		\varphi_j&= \frac{j}{J} \left(1-4\frac{E_J}{J E_M} \right)\left({\varphi_\text{ext}+2\pi f} \right)\mod2\pi . 
	\end{aligned} 
\end{equation}
This correction is small as long as $E_M\ll E_J/J$, establishing $E_J/J$ as the critical energy scale for the validity of our minimization.}

\com{
Crucially, let us note that because of the sinus having period $4\pi$ in Eq.~\eqref{eq:ForMinimazation001}, for $\varphi_J$ we also find extrema close to $\varphi_J\approx \varphi_{\text{ext}}+2\pi-\delta \varphi_J(f+1)$ in addition to the above discussed $\varphi_J\approx \varphi_{\text{ext}}+\delta \varphi_J(f)$. However -- and this is the decisive point -- these correspond to local \textit{maxima} and not minima. Hence they do not correspond to valid local stable points between which the system could tunnel and they therefore play no role for the low-energy physics we are interested in. Consequently, here QPSs change the the phases $\varphi_j$ for $j\neq 0,J$ between different $f$, without changing the value of $\varphi_J$. This fact prompts us to reject the commonly accepted notion of representing quantum phase slips in the form $e^{\pm i 2\pi \widehat{N}_J}$, at least in the here considered parameter regime. In the next section, we will treat the opposing regime, where such jumps are indeed possible due to $E_M$ being small.}

\com{
\subsection{The non-inductive regime $E_L>E_M$}\label{sec_noninductive_regime}

Let us now discuss the aforementioned energy regime of $E_J / J\ll E_M / 2$. For this purpose, it will be more practical to change to the  node variables, $\theta_j= \varphi_{j}-\varphi_{j-1}$, as introduced above, leading to the potential energy $V$ in the representation given by Eq.~\eqref{eq_V_node}. 

For a large $E_M$, we may obtain a very straightforward intuition for the relevant system dynamics by proceeding step by step, starting from a small chain and gradually increasing $J$. In fact, let us start with the simplest possible system of $J=1$. Here, the potential energy is of the form
\begin{equation}
    V=-E_J\cos(\theta_1)-E_M\cos\left(\frac{\theta_1-\varphi_\text{ext}}{2}\right)\ ,
\end{equation}
and for $E_M\ll E_J$ indeed allows for jumps between different minima being separated by $2\pi$, that is $\theta_1\approx 0$ and $\theta_1\approx 2\pi$.

When now increasing $J$, one sees that for the $4\pi$-periodic $\theta_j$ [see also discussion after Eq.~\eqref{eq_V_node}]
 all combinations of $\theta_j\approx 0,2\pi$ are bona fide minima, as long as $E_J/J\gg E_M$. One may now approximate the impact of the capacitive coupling as jumps between different minima along one coordinate, that is, from $\theta_j\approx 0$ to  $\theta_j\approx 2\pi$ (and vice versa), by the operators $e^{\pm i 2\pi \widehat{M}_j}$. Crucially, these operators look suspiciously like the standard representation of QPSs. Note however, that the $4\pi$-periodic constraint remains valid. That is, the $e^{\pm i 2\pi \widehat{M}_j}$ operator is only a (sloppy) short cut for the jump from $0$ to $2\pi$, and should by no means be understood as a ``decompactification'' of $\theta$.
Within the same logic, we now explicitly understand, that this second regime (large $E_J/J$) cannot exist for integer $q/p$ where $\theta$ must be compact on a $2\pi$-periodic interval.

Thus, including the capacitively induced tunneling, we here find an approximate Hamiltonian of the form
\begin{equation}
	\label{eq:HamJJarrayshuntMJ002}
		H\approx- e_{S}\sum_{j}\cos\left(2\pi \widehat{M}_{j}\right)-E_{M}\cos\left(\frac{\sum_{j=1}\widehat{\theta}_{j}-\varphi_{\text{ext} }}{2}\right) \ .
\end{equation}
Note that we expect that in general $e_S\neq E_S$. Secondly, we observe that there is no longer any quadratic term scaling with $E_J/J$, which is why we call this regime the non-inductive regime. At any rate, this treatment is very similar to the one described in Ref.~\cite{Gurarie_2004}. However, we have to insist yet again on the crucial difference, that it is common in the existing literature, to assume $\theta_j$ (or similarly the node variables) to be noncompact, such that the corresponding charge is not at all quantized. Here, we take care to keep charge quantization, again, due to the Majorana shunt, in half-integer units of the Cooper pair charge. The $4\pi$-periodic constraint on the branch phase variables $\theta_j$ then manifests as additional important constraints on the wave function. Namely,
to solve this Hamiltonian we can use the wave function ansatz 
\begin{widetext}
	\begin{equation}
		\begin{aligned}
			\left|\psi\left(f_{1},\cdots,f_{J}\right)\right\rangle =	\sum_{i_1,...,i_J \in \text{EO}} A_{i_1,...,i_J} \sum_{f_{1}\in i_1}\sum_{f_{2}\in i_2} \cdots \sum_{f_{J}\in i_J} e^{ik_{1}f_{1}} \cdots e^{ik_{J}f_{J}} \left|f_{1}, \cdots , f_{J}\right\rangle  \ ,
		\end{aligned}
	\end{equation} 
\end{widetext}
with $\text{EO}=\left\lbrace \text{even numbers}, \text{odd numbers}  \right\rbrace $.
The aforementioned $4\pi$-periodic constraint now results in the condition that $\forall f_i: \left|\psi\left(f_{1},\cdots,f_i+2,\cdots,f_{J}\right)\right\rangle =\left|\psi\left(f_{1},\cdots,f_i,\cdots,f_{J}\right)\right\rangle $, thus $\Rightarrow e^{i2k_i}\stackrel{!}{=}1$ $ \Rightarrow k_i \in \left\lbrace 0,\pi\right\rbrace $. In order to obtain the resulting eigenspectrum, we may again proceed step by step. The first nontrivial example is for $J=2$, where we can rewrite the Hamiltonian in matrix form
\begin{widetext}
	\begin{equation}
	\label{eq:H2}
		H_2=\left(\begin{array}{cccc}
			-E_{M}\cos\left(\frac{\varphi_{\text{ext}}}{2}\right) & -e_{S}\cos\left(k_{2}\right) & -e_{S}\cos\left(k_{1}\right) & 0 \\
			-e_{S}\cos\left(k_{2}\right) & E_{M}\cos\left(\frac{\varphi_{\text{ext}}}{2}\right) & 0 & -e_{S}\cos\left(k_{1}\right)\\
			-e_{S}\cos\left(k_{1}\right) & 0 & E_{M}\cos\left(\frac{\varphi_{\text{ext}}}{2}\right) & -e_{S}\cos\left(k_{2}\right)\\
			0 & -e_{S}\cos\left(k_{1}\right) & -e_{S}\cos\left(k_{2}\right) & -E_{M}\cos\left(\frac{\varphi_{\text{ext}}}{2}\right)
		\end{array}\right)\ ,
	\end{equation}
\end{widetext}	
where $-e_S\cos\left(k_{i}\right) $ can be set to $\pm e_S$ since $k_i$ can only be $0$ or $\pi$.
Hence, the Hamiltonian can be generalized for arbitrary $J$ via:
\begin{equation}
    \label{eq:HJ_JJarray}
	H_J=\left(\begin{array}{cc}
		H_{J-1}\left(E_M\right) & \pm e_S \mathbb{1}_{2^{J-1}} \\
		\pm e_S \mathbb{1}_{2^{J-1}} & H_{J-1} \left(-E_M\right)
	\end{array}\right) \ ,
\end{equation}
where $\mathbb{1}_{k}$ is a $k$-dimensional identity matrix. Hence, $H_J$ can be evaluated iteratively starting from $H_2$, Eq.~\eqref{eq:H2}, by  nesting the predecessor $H_{J-1}$ into the larger two-by-two block structure.
The resulting eigenvalues of this Hamiltonian are of the form 
\begin{equation}
	E_b = \pm\sqrt{E_M^2 \cos^2\left(\frac{\varphi_{\text{ext}}}{2}\right) +b_n^2 e_S^2}
\end{equation}
where 
\begin{equation}
	b_n=\begin{cases}
		2n & \text{for }J\text{ even}, n\in \mathbb{N}_{\leq J/2} \\
		2n+1 & \text{for }J\text{ odd}, n\in \mathbb{N}_{\leq \left\lceil J/2 \right\rceil }
	\end{cases}\ .
\end{equation}
The degeneracy of these states are given by the binomial coefficient
\begin{equation}
	D\left(n\right)=\begin{cases}
		\frac{1}{2}\left(\begin{array}{c}
            J\\
            J/2
        \end{array}\right) & \text{for }J\text{ even and } n=0 \\
		\left(\begin{array}{c}
            J\\
            \left\lceil J/2 \right\rceil-n
        \end{array}\right) & \text{else }
	\end{cases}\ .
\end{equation}
Overall, we see that the energy spectrum is here of a very comparable form as Eq.~\eqref{eq:EEShuntedMJ1}, except that the density of states for the phase slip contribution (related to the energy $b_n e_S$) is different now due to $D(n)$. Note that said contribution $b_n e_S$ goes from $0$ to $J e_S$. We thus introduce a macroscopic energy for the quantum phase slips, defined as $E_S=J e_S$, in order to be able to compare with the model discussed in the main text and the results obtained in Sec.~\ref{sec_DOS} above.

\begin{figure}[htbp]
    \begin{center}
        \includegraphics[width=0.3\textwidth]{EvolutionDOS002.png}
    \end{center}
    \caption{Evolution of the density of states (DOS) for Hamiltonians of the from \eqref{eq:HJ_JJarray}, with (a) $J=16$, (b) $J=86$ and (c) $J=186$. }
	\label{fig:EvolutionDOS}
\end{figure}

In order to evaluate the DOS, we first compute the contribution for a given $n$
\begin{equation}
    \rho_n(E)=\frac{1}{\pi }\text{Re}\left[\frac{\sqrt{E^2}}{\sqrt{E^{2}-b_n^{2}e_{S}^{2}}\sqrt{E_{M}^{2}-E^{2}+b_n^{2}e_{S}^{2}}}\right]
\end{equation}
and subsequently take into account the degeneracy $D(n)$ for each $n$, to arrive at
\begin{equation}\label{eq_DOS_bn}
    \rho\left(E\right)=\frac{1}{D_\text{tot}}\sum_{n=0}^{\left\lceil J/2 \right\rceil} D\left(n\right)\rho_n(E) \ ,
\end{equation}
with $D_\text{tot}=\sum_{n=0}^{\lceil J/2\rceil} D(n)$. This density of states is normalized to $\int dE\rho(E)=1$, just like the previous results in Eqs.~\eqref{eq_DOS_analytic} and~\eqref{eq_DOS_analytic_MJ}. Note that the above sum will give rise to quite a large number of van Hove singularities, for each summand at $E=\pm b_n eS$ and $E=\pm\sqrt{E_M^2+b_n^2 e_S^2}$. In order to be able to provide a smooth transition to the continuum limit $J\rightarrow\infty$, we introduce a level broadening, which can be done by a simple imaginary shift of the energy, $\rho_n(E)\rightarrow \rho_n(E+i\delta)$.

The density of states for three different numbers of JJs is depicted in Fig.~\ref{fig:EvolutionDOS}, for $E_M=E_S$, by keeping $E_S=e_S/J$ and $\delta$ constant.
Notice that as $J$ gets bigger, the level broadening becomes bigger relative to the the distance between the energies $b_n e_S$, such that $\rho(E)$ smoothens out. Thus, for large $J$ the DOS near $0$ starts to resemble a Dirac cone, like in Fig.~\ref{fig:PhaseSlipJJMJ}c. However, the similarity between the DOS of Eq.~\eqref{eq_DOS_bn} and Eq.~\eqref{eq_DOS_analytic_MJ} is only of qualitative nature. For instance, in Fig.~\ref{fig:EvolutionDOS} the decline after the peaks is steeper than in Fig.~\ref{fig:PhaseSlipJJMJ}c.
}

\subsection{Local versus non-local charge definition}

As we discussed in the main text, we believe that neglecting charge quantization in QPS junctions is intimately connected to neglecting the phase dependence of the QPS basis $|f\rangle$. Here we want to elucidate this point a bit more explicitly. This discussion is inspired by the findings of Ref.~\cite{Roman2021}.

\com{
To begin, let us point out, that of course, this discussion is only relevant for the inductive regime (see Sec.~\ref{sec_inductive_regime}), as otherwise there is no inductive energy $E_L$, see the Hamiltonian given in
Eq.~\eqref{eq_H_new_QPS_alone}. }

As already stated above, we can think of the minima given in Eq.~\eqref{eq:MinAnsatzJJarray} as the reference points around which the wave function $|f\rangle$ is centered. Since the minima in Eq.~\eqref{eq:MinAnsatzJJarray} depend on the total phase difference $\varphi$ between the two contacts linked by the QPS junction, the basis of $\widehat{f}$ is likewise $\varphi$-dependent, $|f\rangle_{\varphi}$. As pointed out in the main text, this phase dependence of the minima can be removed by a unitary transformation, $U\left(\varphi\right)=\sum_f \vert \widetilde{f} \rangle\left\langle f \right\vert_\varphi$.

In the space of the individual phases $\varphi_j$, this unitary is simply a translation operator,
\begin{equation}
    U=\prod_{j=1}^{J-1} e^{-i \frac{j}{J}\varphi \widehat{N}_j}\ .
\end{equation}
This realization allows us now to formulate the two different charges. Originally, $\varphi$ is only defined on the right-most contact. Its conjugate charge, here denoted as $N_q$ is guaranteed to have integer eigenvalues in the finite element model given by the Hamiltonian in Eqs.~\eqref{eq_T_of_H} and~\eqref{eq:HamJJarrayshuntMJ}, because it is assigned to a sharply defined region, see also the arguments in Ref.~\cite{Roman2021}. In fact, $N_q$ is the charge that should enter the Hamiltonian, if we embed the QPS junction into a larger circuit, e.g., shunting it with a Josephson junction element or simply a capacitive element, as in Fig.~\ref{fig:QPS_standard_treatment_start}b. The charge emerging from the unitary transformation on the other hand is given by
\begin{equation}
    \widehat{N}=\widehat{N}_q+i U^\dagger \partial_{\varphi} U = \widehat{N}_q +\sum_{j=1}^{J-1}\frac{j}{J}\widehat{N}_j \ .
\end{equation}
We now see very explicitly, that while each local charge within the array, $\widehat{N}_j$, has again integer eigenvalues, the transformed charge $\widehat{N}$ does not, due to the prefactor $j/J<1$ (obviously, a rational number, which, in the continuum limit $J\rightarrow \infty$ approaches an irrational value). 
As a matter of fact, the transformed charge $\widehat{N}$ can be understood as a ``fuzzy'' measurement of the charge, with a support falling off linearly along the JJ-array, see Fig~\ref{fig:PhaseSlipExpression2}c in the main text. Importantly, as we argue in the main text, when removing the high-energy state, we can locally flatten the basis. That is, due to the low-energy approximation, the originally quantized charge $N_q$ is removed from the dynamics, and the fuzzy charge $N$ takes its place. Nonetheless, the compactness of the phase $\varphi$ remains important, as it fixes the total number of available low-energy states (that is, the number of local minima).

\section{Solution for two QPS} 
In principle the compactness condition in the main text (that is, the periodic constraints in Eqs.~\eqref{eq_boundary_psi} and~\eqref{eq_psi_boundary_2QPS}, for circuits including one or more QPS junctions) in conjunction with the standard node-flux are sufficient to compute the correct energy eigenspectra. The (superfluous) larger Hilbert space is carried around for the full calculation of the eigenvectors and eigenenergies of the Schr\"odinger equation, and can be reduced by imposing the periodic constraints. However, our new treatment also allows to calculate right from the start with the reduced Hilbert space, allowing for significant reduction of the computational effort.

As an example we will discuss here a system consisting of two QPS-junctions connected in series with a gate voltage in between (Fig. \ref{fig:QPS_junctions_standard_description_AB_AC}), with the Hamiltonian given in the main text in Eq.~\eqref{eq_H_2QPS_compact}. This Hamiltonian can be readily solved numerically by discretizing $\varphi$-space and imposing the boundary condition Eq.~\eqref{eq_psi_boundary_2QPS} on the discrete lattice.
\com{
\rr{\textbf{[Why is there a new Hamiltonian? And what do you mean with "two meaningful charges"? We already have the right Hamiltonian in the main text! I think this part (everything marked in red) can just go.]} \sout{Here we have two meaningful charges $N$ and $\widetilde{N}$ (canonically conjugate to the phases $\varphi$ and $\widetilde{\varphi}$ respectively)  as well as the gate charge $N_g$. This system will obey the Hamiltonian}
\begin{equation}
    \label{eq:2QPS_H}
    \begin{aligned}
    H=&-E_{S_1} \cos\lr{S_1} + E_{L_1} \lr{\varphi-2\pi f_1}^2\\
    &-E_{S_2} \cos\lr{S_2} + E_{L_2} \lr{\widetilde{\varphi}-\varphi-2\pi f_2}^2\\
    &+E_{C_g} \lr{N+N_g}^2 \ ,
    \end{aligned}
\end{equation}
\sout{where $\lrB{S_i,f_j}=i \delta_{ij}$.} }

\rr{\textbf{Did Mooij and Nazarov talk about effective Josephson junctions? I don't remember that...} \sout{Analogous to Ref.~\cite{Mooij_2006}}}} There is a particularly simple regime, for $E_{Si}> E_{Li}$ ($i=1,2$) where we can approximate the effect of the two QPS junctions as effective Josephson junction with a Josephson energy given in Eq.~\eqref{eq_EJeff}.
Thus we find a new effective Hamiltonian of the form
\begin{equation}
    \label{eq:2QPS_H_Eff}
    \begin{split}
    H_{\text{eff}} = - E_{J,\text{eff}_{1}} \cos\lr{\wphi} - E_{J,\text{eff}_{2}}\cos\lr{\wphi-\phi_\text{ext}} \\+E_{C} \lr{\wN+N_g}^2\ .
    \end{split}
\end{equation} 
This is straightforwardly mapped onto a regular charge qubit Hamiltonian, whose solutions are known, via the Mathieu functions~
\cite{Cottet_2002}. These solutions give rise to the energy spectrum for the compact node-flux approach, see third panel in Fig.~\ref{fig:standartVSloopVSCompact}. The corresponding standard node-flux and loop-charge results (first and second panels) can be obtained by replacing $N_g$, respectively, $\phi_\text{ext}$ by Bloch vectors $k$, and plotting the resulting energy band widths.

\bibliography{bibliography,biblio}